\documentclass[showpacs,preprintnumbers,a4paper,acs,12pt,longbibliography]{revtex4-1}

\usepackage{epsfig}
\newcommand{\be}{\begin{equation}}
\newcommand{\ee}{\end{equation}}
\newcommand{\bd}{\begin{displaymath}}
\newcommand{\ed}{\end{displaymath}}
\newcommand{\bea}{\begin{eqnarray}}
\newcommand{\eea}{\end{eqnarray}}

\begin{document}

\title{ A novel nanoporous graphite based on graphynes: first principles
  structure and carbon dioxide preferential physisorption  }

\author{Massimiliano Bartolomei\footnote{Corresponding author, e-mail:maxbart@iff.csic.es},
\\Instituto de F\'{\i}sica Fundamental, Consejo Superior de Investigaciones Cient\'{\i}ficas (IFF-CSIC), Serrano 123,
28006 Madrid, Spain,
\\Giacomo Giorgi
\\Dipartimento di Ingegneria Civile ed Ambientale (DICA),\\
 University of Perugia,\\ Via G. Duranti 93, I-06125\\ 
Perugia, Italy.}


\begin{abstract}
Ubiquitous graphene is a stricly 2D material representing an ideal adsorbing platform due to its large
specific surface area as well as its mechanical strength and resistance to both
thermal and chemical stresses. However, graphene as a bulk material has the
tendency to form irreversible agglomerates leading to 3D graphitic structures
with a significant decrease of the area available for adsorption and no room for gas intercalation. 
In this paper a novel nanoporous graphite formed by graphtriyne sheets is
introduced: its 3D structure is theoretically assessed by means of
electronic structure and molecular dynamics computations within the DFT level
of theory.
It is found that the novel layered carbon allotrope is almost as compact as pristine
graphite but the inherent porosity of the 2D graphyne sheets and its relative stacking leads to
nanochannels that cross the material and whose sub-nanometer size could allow the
diffusion and storage of gas species. 
A molecular prototype of the nanochannel is used to accurately determine
first principles adsorption energies and enthalpies  for CO$_2$, N$_2$,
H$_2$O and H$_2$ within the pores. The proposed porous graphite presents no significant
barrier for gas diffusion and it shows a high propensity for CO$_2$
physisorption with respect to the other relevant components in both pre- and  post-combustion gas streams.

\end{abstract}

\date{\today}

\maketitle


\section{Introduction}

The use of advanced porous materials for gas storage has recently received
wide attention since they can help to provide solutions to growing concerns
 related to the anthropogenic carbon dioxide (CO$_2$) emissions which, in
 turn, are expected to rise considerably\cite{urgenPNAS:12,natclimchange:13} if fossil fuels consumption will continue 
 to dominate the energy scene.
 
In fact, porous materials could conveniently and  efficiently
reduce the carbon content in gas mixtures by exploiting the selective
physical adsorption of CO$_2$ at the adsorbent's surface  or within its pore network.




Activated carbons and zeolites are traditionally used to this scope  and, more
recently, promising adsorbents such as metal-organic frameworks
(MOF)\cite{MRS:13} have attracted much attention. 

The crystalline structure of MOF implies noticeable advantages with respect to traditional porous
materials: in fact, their regular and
ordered porosity, together with tunable size and shape of the openings, has
led to very high performances in gas uptake
capacities\cite{ChemSocRev:12,revco2:14}. However, their thermal stability as well as
the vulnerability of the metal containing clusters to ligand substitution by water or other nucleophiles 
still represent not negligible shortcomings. 

Therefore, investigations on alternative porous materials are highly advisable
and, specifically, on those based on carbon since they provide a large 
 specific surface area together with a low sensitivity to humidity as well as a low weight. 
As an example, very recently novel porous carbons have been synthesized\cite{porouscarb:14} by means of
a high temperature carbonization process of different MOF structures: very high CO$_2$ uptake capacities have been found 
 which even improve those of the metal-organic precursors.   

Ubiquitous graphene represents in principle the ultimate adsorbing carbon
platform\cite{graphadsrev:15} due to its high surface area and mechanical strength, excellent
thermal conductivity and good chemical stability,
 but the related gas uptake capacity is limited by its tendency to
  form irreversible agglomerates and intrinsic lack of
 pores. As a matter of fact bulk graphene tends to cluster in graphite platelets due to
 the large van der Waals forces between the large and planar basal planes and
 this leads to a significant decrease of the available surface area.
  Multi-layered graphenes or graphites could indeed lead to an enhanced gas storage
 after interlayer spacing expansion and/or activation. As an example, a larger
 separation between graphene-like planes has been obtained\cite{Graphoxi:10} from graphene oxide layers
 connected by molecular pillars but the related measured gas uptakes have been
 found quite lower than those predicted by the simulations and in any case not
 comparable with the best values reached with MOF
 materials\cite{ChemSocRev:12,revco2:14}. 
Nevertheless, activated graphene-derived powders and structures heve been recently
obtained\cite{carbonnanostruc:15,activgraph:16} from exfoliated graphite
oxide:  these carbon materials have indeed shown exceptional gas adsorption
properties which are comparable to those for MOFs having similar surface area.

 The introduction of nanopores into graphene sheets is another effective
 possibility for improving their gas adsorption performances.
In fact, in the last years two-dimensional (2D) materials similar to graphene but with
regular and uniformly distributed subnanometer pores have
been synthesized in large area films\cite{Bieri:09,graphd-chemcomm:2010} by means of ``bottom up'' approaches.
 Among them $\gamma$-graphynes\cite{narita:98,graphd_review:2014}, which are new 2D carbon allotropes formed by sp-sp$^2$
hybridized carbon atoms, are of particular interest and they feature triangular pores of sub-nanometer size.
In fact, their recent synthesis\cite{graphd-chemcomm:2010,graphdsynt:15} has triggered important theoretical
studies devoted to their application as effective single-layer membranes for gas separation and 
water filtration technologies\cite{nanoscalemitbis:2012,nanoscalemit:2013,jpclours:2014,jpccours:2014,heliumisotop:15}. 

In addition to their use as sieving platforms, graphynes could be also
promoted as adsorbents since the 2D sheets can be used as building blocks to
construct novel graphyne-based 3D porous structures\cite{bitri:2012,bulk-graphyne:2013,bulk-graphd:2013}.

As a matter of fact, we have recently shown \cite{carbonour:2015} by means of
a theoretical investigation that multi-layer $\gamma$-graphynes are more
suited than graphene for the physical adsorption of molecular hydrogen
(H$_2$). In particular, those based on graphtriyne layers (see Fig. 1), which are
formed by phenyl rings fully interconnected by chains of three conjugated C-C triple bonds, can allow
both H$_2$ intercalation and its diffusion through the carbon material and the
computed adsorption energies inside the pores are almost doubling the related
estimations\cite{Vidali:91,dft-cc:10,dft-vdw:14} for the adsorption on pristine graphene.
 The size of the involved triangular pores ( pore width below 0.7 nm )
 suggests they could be also effective to host molecules larger than H$_2$ and 
 therefore their affinity for CO$_2$ physisorption can be postulated and it is
 worth to be further investigated. 


In this work we first aim to correctly determine the 3D structure of a novel nanoporous graphite
 composed of stacked graphtriyne layers. Then, its capability for the 
 preferential adsorption of CO$_2$ with respect to other
 gases such as nitrogen,  water and hydrogen is 
 theoretically addressed.
 




\section{Computational Methods}
\label{sec.2}

The calculations for the graphtriyne 3D structure have been obtained by
means of density-functional theory (DFT), as implemented in the VASP code
\cite{VASP:96}, within the generalized gradient approximation (GGA) of Perdew, Burke, and Ernzerhof (PBE)\cite{pbe:96}.
The Bl\"{o}chl all-electron projector-augmented wave (PAW) method
\cite{PAW1,PAW2}, with an energy cutoff of 750 eV and a 2s$^2$ 2p$^2$ electron valence potential for carbon, has been
employed. A periodic model has been first considered for
the graphtriyne isolated monolayer with an initial supercell (see Fig. 1) constituted by 48 C
atoms  with in-plane lattice parameter {\it a} =  12.035 \AA\,  and  {\it b} =
20.841 \AA, and with a sufficient vacuum amount to avoid spurious interaction
between layers along the direction normal to the {\it ab} plane.
 According to the large size of the parameters, we have sampled the Brillouin
 zone (BZ) with a 5x4x1 Gamma centred mesh. 
Then a periodic model for the graphtriyne isolated bilayer with 48 atoms per layer (96 C
atoms) has been taken into account in order to study the bilayer interaction energy.  
Keeping the supercell lattice parameters fixed to the initially optimized
values, by means of single point calculations, we have progressively slided the top
layer upon the bottom one along both {\it a} and {\it b} directions, respectively(see Fig. 1), at a
fixed interlayer distance $R$= 3.45 \AA\, which has been previously found\cite{carbonour:2015} to be the
optimal one for the bilayer interaction.

 The computed energies have been corrected by two- and three-body dispersion contributions including the Becke-Johnson
 damping scheme\cite{Grimme:11} and an Axilrod-Teller-Muto three-body term\cite{Grimme:14},  as
 implemented in the $dftd3$ program of Grimme et al.\cite{Grimme:10}. 


Moreover, in order to better assess the actual stability of the multilayer
material we have performed ab initio finite temperature molecular
dynamics (AIMD) simulations by considering an isolated trilayer prototype: the time step 
 has been taken as 2.5 fs and atomic velocities have been renormalized to the
 temperature set at T= 77 K at every 40 time steps. The 12 ps MD simulations have been performed with a sparser 3x2x1 Gamma centred sampling of the BZ.

The electronic structure calculations for the adsorption energy of CO$_2$,
N$_2$ and H$_2$O within the pores have been carried out at the
 ``coupled'' supermolecular second-order M{\o}ller-Plesset perturbation theory
(MP2C)\cite{mp2c} level of theory by using the Molpro2012.1
package\cite{MOLPRO}.
Our choice to use the MP2C approach relies on its capability to provide reliable estimations for weakly bound systems
such as rare gas--fullerene\cite{mp2c-full} and  -coronene\cite{grapheneours:2013}
as well as molecule--graphynes' pores\cite{jpclours:2014,carbonour:2015}, at an affordable computational cost.
For the graphtriyne pore prototype we have considered the following bond
lengths\cite{mech-graphdiyne:2012}, which are those obtained from the periodic
DFT calculations: 
1.431 \AA\, for the aromatic C-C, 1.231 \AA\,
for triple C-C, 1.337 \AA\, for the single C-C between two triple C-C bonds,  
1.395 \AA\, for the single C-C connecting aromatic and triple C-C bonds. 
The size of the C-H ends of the pore prototype is
1.090 \AA\, while C-O, N-N and O-H bond lengths are 1.162, 1.100 and 0.957 \AA,
respectively.
In the case of H$_2$O, the HOH angle is considered to be 104.5 degrees.
 The aug-cc-pVTZ\cite{Dunning} basis set has been employed for the pore structures, 
while the aug-cc-pVQZ\cite{Dunning} basis has been used for the interacting
molecules. 
In the MP2C computations all considered molecular structures are treated as rigid bodies: the atoms composing the
investigated  graphtriyne prototype are frozen in their initial positions and the
molecular configuration of CO$_2$ (and N$_2$ and H$_2$O) is not allowed to relax during the calculations.  
The interaction energies have been further corrected for the basis set
superposition error by the counterpoise method of Boys and
Bernardi\cite{Boys:70}. 
The zero point vibrational energy corrections to be added to the binding energies in order to
provide an estimation of the adsorption enthalpies have been obtained by
calculating the vibrational harmonic frequencies for the interaction of the interested
molecule placed within an inner pore of the prototype of the multi-layer (see
Fig. 4). The latter have been considered as a rigid substrate and
the calculations have been performed at the DFT level of theory by
exploiting the PBE functional along with the cc-pVTZ\cite{Dunning} basis set
and the latest dispersion contribution correction of Grimme\cite{Grimme:11} 
 as implemented in the Gaussian 09 code\cite{g09}. We have checked that
 this level of theory provide a good estimation of the involved interaction 
if compared with the reference MP2C calculations as shown in Fig. S1, 
where the case of the molecular interaction with a single
 pore is reported.

\section{Results and Discussion}
 
\subsection{Porous graphite structure}

The novel graphite we propose is composed of stacked graphtriyne
sheets and the accurate assessment of its 3D structure represents a
challenging task since the strong interlayer dispersion forces leading to the compact
carbon material must be correctly taken into account.
To do that we rely on the approach recently introduced by Brandenburg et
al. \cite{Brandenburg:2013} which provides reliable dispersion
corrections to periodic DFT calculations: as a matter of fact its application guaranteed accurate
and reference values\cite{Brandenburg:2013}
  for both exfoliation energy and interlayer separation of pristine graphite.
In our previous work\cite{carbonour:2015} the same approach has been used to
determine the interlayer equilibrium
 distance of the graphtriyne bilayer for a specific stacking
configuration: we have found it to be
 3.45 \AA, which is just 0.1 \AA\, larger than that for pristine graphite.
Here, we determine in details the energy features of the bilayer relative
configuration in order to correctly determine
 its optimal stacking: this represents a critical
point since it could occur that the pore of one layer is partially or totally
obstructed by the phenyl ring of the adjacent layer and
 thus leading to a difficult or impeded diffusion of atoms and
molecules through the multi-layer.
Accordingly, we have performed DFT calculations for the bilayer interaction energy at a fixed interlayer distance
(equals to 3.45 \AA) and as a function of the shift of one layer upon
the other along two different directions.
 In the left part of  Fig. 1 we report the optimized structure of the 2D
graphtriyne monolayer used
to perform the periodic DFT calculations of the bilayer interaction: the in-plane lattice parameters are also reported and throughout the calculations they have been considered  fixed as well as the in-plane distances between carbon atoms, that is both carbon
 sheets have been considered as rigid bodies.
 In the right part of  Fig. 1 we show the energy profiles originated
from a series of single point calculations
 related to the parallel displacement of one layer upon the other with
respect to a AA-stacked bilayer corresponding
 to the zero of the abscissae: we can see that for both {\it a} and
{\it b} directions an almost coincident minimum around 1.55 \AA\, is
 found and the corresponding bilayer configurations, labelled as AB-1
and AB-2, are also represented.
The AB-1 configuration recalls that of the typical Bernal stacking of
pristine graphite, while the AB-2 one is similar but with a comparable shift in
the in-plane perpendicular direction.
Similar minima (not reported in Fig.1) have been also found for displacements along 
 directions others than {\it a} and {\it b}. 
The binding energies are -1454.2 and 1452.9 meV for the
AB-1 and AB-2 configurations, respectively, which correspond to about 30.3
meV/atom, to be compared to 42.9 meV/atom as estimated at same level of theory 
 for a graphene bilayer\cite{Brandenburg:2013}. Clearly, for a graphtriyne
 bilayer a decrease of about 30\% in the binding energy is found with
 respect to the graphene counterpart and this can be understood considering 
 that a graphtriyne plane contains a lower
density of carbon atoms: in fact, a graphtriyne layer can be thought as the
result of replacing one-third of the carbon-carbon bonds in graphene with triple
conjugated acetylenic linkages. Also, from Fig. 1 it can be evidenced 
 that further displacements from the main minimimum at 1.55 \AA\, can lead to another less stable
 one around 3.5-4.0 \AA\, and the barrier to be overcome is not large being 
 about 125.0 meV ($\sim$ 2.6 meV/atom). 
At non-zero temperatures, these energy features could lead to a not well
defined structure for the multi-layer with large in-plane and interlayer displacements.
Therefore it is worth further investigating the stability of the 3D structure of
  by performing DFT/AIMD simulations.  To do that we have
  considered as a starting point two different trilayers in ABA-1 and ABA-2 stackings, 
 which are related to the AB-1 and AB-2 minimum geometries found for the bilayer.
 Then molecular dynamics simulations are run at constant temperature  T= 77K
 (typical in gas adsorption experiments)
 for about 12.5 ps and we have found that both starting points eventually
 lead to a similar ABA structure. Once the system equilibration has been
 reached (after about 9 ps),
 radial distributions of both the interlayer
 distance and inner plane layer displacement have been computed and they are
 reported in Fig. 2.
As for the interlayer distance we have considered the distances between corresponding carbon
atoms constituting the phenyl rings lying on the outer layers since during the
simulations their relative position on the {\it ab} plane remains almost constant.
As for the inner layer shift we have taken into account the relative position on the  {\it ab} plane 
of the carbon atoms constituting the phenyl rings with respect to the
corresponding ones on the external sheets.
 
Our results show that the interlayer distance between the outermost layers ranges
between 6.7 and 7.1 \AA, an interval whose middle point coincides with the double of the equilibrium
distance (3.45 \AA) we previously found\cite{carbonour:2015} for the bilayer.
In the case of the inner layer shift we have found that the most prominent
probability is  between 1.3 and 1.9 \AA, that is around the main minima shown in
Fig. 1. These results clearly demonstrate that, despite their weaker
interlayer binding with respect to pristine graphite, graphtriyne layers tend
to aggregate and to maintain a stable multi-layer structure at finite
temperatures: this structure is of a ABA-1 type (see Fig. 1) and the
displacement of the inner layer with respect to outer ones is limited. 
 The latter ensures that in the bulk material the free sliding of one layer upon the others
 is impeded and, more importantly, that the pore of one layer is sligthly
 displaced from the corresponding ones on the  adjacent layers leading to a
 sort of nano-channels that cross throughout the multi-layer.
 These nano-channels could be exploited for gas diffusion and adsorption and 
 the in the following we analyze the capability of the novel nano-porous 
 graphite to selectively adsorb CO$_2$ molecules.


\subsection{CO$_2$ selective adsorption}
First, the interaction between the gas molecule and a graphtriyne single pore is
accurately determined. In order to do it, we have considered a molecular precursor of 
 the pore represented by the octadecadehydrotribenzo[24]annulene (Fig. 3, top panel), together with the most stable in-pore configuration of the considered gas molecules, that is CO$_2$, N$_2$, H$_2$O and H$_2$.
In the case of N$_2$ and CO$_2$ the optimal geometry is that perpendicular to
the pore plane with the center mass of the molecule lying in the geometrical
center of the opening; for H$_2$O and H$_2$ instead the most favourable geometries are
co-planar to the pore as clearly reported in the figure. 
In the lower part of Fig. 3 the corresponding interaction curves obtained at
the MP2C level of theory are depicted as a function of the gas molecule distance from the
center of the pore. The molecule configurations are kept frozen during the
calculations and are those represented in the upper part of Fig. 3.
It can be seen that in all cases the minimum is located just inside the
pore, that is the size of the opening is large enough to host the considered molecular
species.
Moreover, the largest binding energy corresponds to CO$_2$ as it could be
expected considering that the involved interaction is mostly determined 
 by the van der Waals contribution being the considered pore a neutral and 
  non-polar substrate: in such cases  both the dispersion attraction and size
  repulsion are mainly governed by the molecular polarizability and that of CO$_2$\cite{Denbigh} is
  the largest among the selected species. 
The binding energy of CO$_2$ adsorbed within a graphtriyne pore is about 200
meV and it also slightly larger than the best empirical (178 meV\cite{Vidali:91}) and
theoretical (198 meV\cite{dft-cc:10}) estimates for the physisorption on a
graphene plane. A similar trend can also be found for N$_2$, H$_2$O and H$_2$.
The results of Fig. 3 show a remarkable preferential adsorption of CO$_2$ over H$_2$ within
 an isolated graphtriyne pore,  which suggests a possible application for carbon capture in pre-combustion
 gas streams; on the contrary the CO$_2$ binding energy is not sufficiently far from that of H$_2$O
 and N$_2$, and therefore the capability of carbon capture and separation does
 not seem likely in post-combustion gas mixtures.
However, one could expect that the separation between binding energies
related to the different species will increase if the adsorption within 
the pore lying on an inner layer of the multi-layered graphtriyne is taken into account.

In order to verify this hypothesis we have built up a molecular
prototype\cite{carbonour:2015} 
of the nano-channel by considering three parallel graphtriyne pores arranged 
 in a ABA-1 stacking as depicted in the upper part of Fig. 4: by using the 
  geometry parameter determined in the previous section we have placed
 the inner pore with a lateral shift of 1.6\AA\, with respect the outer ones, while the adjacent layers are separated by 3.45 \AA.

In Fig. 4 the positions
 of the specific molecule considered for the calculations
 are also reported and they are indicated in blue as A, B,
 C, B' and A'. In particular, the A, A' and C sites correspond to in-pore configurations
 in which the molecule is right in the pore geometric center. B and B'
 equivalent sites correspond instead to the molecule intercalation:
 the molecule center of mass lies right in the direction (dashed line) joining
 the geometric centers of adjacent pores and it is placed  at 1.725 \AA\, (half the interlayer
 distance) from the closest layers.
In the lower part of Fig. 4 the adsorption energies obtained at the MP2C level
are reported for the five adsorption sites
 and each value corresponds to the sum of three contributions related to the specific molecule interacting with each of the
 graphtriyne pore. 
The intention is to provide the evolution of the adsorption energy
 as the molecule crosses the nanoporous graphite.   

For the in-pore A, A', and C sites the obtained interaction
energies are significantly larger than those for the single pore case (see Fig. 3 ): clearly the largest interaction corresponds to the C case (that for the 
 specific molecule inside the intermediate pore) and the related adsorption
 energies for CO$_2$, N$_2$, and H$_2$ are more than 50\% larger than those for
 the single pore. 
The interaction improvement is less evident in the case of H$_2$O whose 
adsorption energy increases of about 40\%: this could be explained considering
that polarization contributions to the global interaction due to the water
dipole moment are less effective for the interaction with the outer and indeed farther pores.

As for the B and B' intercalation sites, the interaction energies are more
attractive than those for the more external in-pore A and A' locations,  even if less than those for the C position.


In general the results reported in Fig. 4 suggest several conclusions.
The first point is that no significant hindrance is present for molecule penetration
across the multi-layer:in fact, when passing from a pore to the adjacent one along the
$z$ direction no relevant barrier  (that is the energy difference between the C and B sites) is observed.
 This means that the out-of-plane gas diffusion through the novel porous graphite could be possible. 

Interestingly, in the case of CO$_2$ the energy difference between the A and B
sites is more pronounced (more than 50 meV) than that for the other species suggesting that once
the molecule is hosted inside the material its release would be more difficult.


The second point to highlight is that a quite favourable 
binding  energy (about 310 meV) is found for CO$_2$ within the inner pore and
also that it is about 100 meV larger than those for 
H$_2$O and N$_2$.  
In order to better assess the preferential adsorption of CO$_2$ an estimation
of the adsorption enthalpies, which represent more realistic magnitudes for
the single molecule--substrate interactions here investigated,
 have been also computed: to do that zero point vibrational energy corrections
 have been calculated at the DFT level of theory 
 and added to the binding energies 
(All the results 
are summarized in Table 1).
It is straightforward that the adsorption
enthalpy of CO$_2$ is significantly larger than those of the other considered
species: the energy difference ranges from about 235 meV for H$_2$ to 110 meV for
H$_2$O and N$_2$.
These data suggest that a preferential adsorption of CO$_2$
over H$_2$, H$_2$O, and N$_2$ could be expected within the pores of the novel
nanoporous graphite. These feautures, together with a moderate adsorption
enthalpy for CO$_2$ (about 300 meV), which entails a strong physisorption,
promote the new carbon layered material as an efficient adsorbing medium not
only for pre-combustion capture processes but also for those involved in post-combustion
where wet CO$_2$/N$_2$ mixtures have to be treated.

As an example, on the basis of stoichiometric considerations and assuming that one 
 CO$_2$ molecules could be hosted in each pore we can provide an  estimation
 of the related gravimetric storage capacity of about 5.3 mmol\,g$^{-1}$ (23.4 wt\%), which 
 is in the range or even higher than those for the best carbon adsorbing materials to date\cite{revco2:14,porouscarb:14,activgraph:16} at low pressures ($\sim$  1 bar), which are those of interest in post-combustion flue gas.
 Indeed, the uniformly distributed sub-nanometer triangular pores featuring the novel graphite are expected 
 to provide good uptake performances at atmospheric and lower pressures since it has been
 found\cite{CO2porediamter} that at such conditions the CO$_2$ sorption is mostly determined
 by pores with diameter smaller than 0.8 nm.
 
\section{Conclusions}

In summary, by means of electronic structure and molecular dynamics computations,
 we have shown that
graphtriyne layers tend to associate themselves at finite
 temperatures to form stable layered structures representing a new kind of
 porous graphite. The optimal structures have
 been identified and they correspond to ABA-like stackings with 
an average interlayer separation of about 3.45 \AA\,; moreover we have shown
that the displacement of one layer with respect the adjacent ones is limited
($\sim$ 1.6 \AA) allowing the spatial connection between corresponding
pores lying on different sheets and leading to a sort of nanochannels which 
 perpendicularly cross the 3D material.
We have shown that these nanochannels are large enough to host light gases
such as CO$_2$, N$_2$, H$_2$O and H$_2$ allowing also their diffusion. The corresponding adsorption
energies and enthalpies have been computed: a strong physisorption is found
for CO$_2$ and a significant selectivity is expected over the other considered
species which could be exploited in both pre- and post-combustion carbon
capture techniques.

The novel porous graphite we propose could be considered as a promising 
alternative to more traditional adsorbing materials based on carbon;
as most of them it is hydrophobic, chemically inert and thermally stable but
with the advantage of a very compact cristallyne structure with regular pores uniformly distributed and of 
sub-nanometer size, which are especially suited to host light gas species.

\section*{Acknowledgments}

The work has been funded by the Spanish grant FIS2013-48275-C2-1-P. 
Allocation of computing time by CESGA (Spain) 
is also  acknowledged.

\section*{Supplementary Material}
  Intermolecular potentials obtained at the DFT level of theory for the
  molecule--graphtriyne pore interaction are reported in an additional figure 
  where they are compared with corresponding MP2C computations.
  This information is available free of charge via the Internet.


\begin{thebibliography}{45}%
\makeatletter
\providecommand \@ifxundefined [1]{%
 \@ifx{#1\undefined}
}%
\providecommand \@ifnum [1]{%
 \ifnum #1\expandafter \@firstoftwo
 \else \expandafter \@secondoftwo
 \fi
}%
\providecommand \@ifx [1]{%
 \ifx #1\expandafter \@firstoftwo
 \else \expandafter \@secondoftwo
 \fi
}%
\providecommand \natexlab [1]{#1}%
\providecommand \enquote  [1]{``#1''}%
\providecommand \bibnamefont  [1]{#1}%
\providecommand \bibfnamefont [1]{#1}%
\providecommand \citenamefont [1]{#1}%
\providecommand \href@noop [0]{\@secondoftwo}%
\providecommand \href [0]{\begingroup \@sanitize@url \@href}%
\providecommand \@href[1]{\@@startlink{#1}\@@href}%
\providecommand \@@href[1]{\endgroup#1\@@endlink}%
\providecommand \@sanitize@url [0]{\catcode `\\12\catcode `\$12\catcode
  `\&12\catcode `\#12\catcode `\^12\catcode `\_12\catcode `\%12\relax}%
\providecommand \@@startlink[1]{}%
\providecommand \@@endlink[0]{}%
\providecommand \url  [0]{\begingroup\@sanitize@url \@url }%
\providecommand \@url [1]{\endgroup\@href {#1}{\urlprefix }}%
\providecommand \urlprefix  [0]{URL }%
\providecommand \Eprint [0]{\href }%
\providecommand \doibase [0]{http://dx.doi.org/}%
\providecommand \selectlanguage [0]{\@gobble}%
\providecommand \bibinfo  [0]{\@secondoftwo}%
\providecommand \bibfield  [0]{\@secondoftwo}%
\providecommand \translation [1]{[#1]}%
\providecommand \BibitemOpen [0]{}%
\providecommand \bibitemStop [0]{}%
\providecommand \bibitemNoStop [0]{.\EOS\space}%
\providecommand \EOS [0]{\spacefactor3000\relax}%
\providecommand \BibitemShut  [1]{\csname bibitem#1\endcsname}%
\let\auto@bib@innerbib\@empty
\bibitem [{\citenamefont {Lackner}\ \emph {et~al.}(2012)\citenamefont
  {Lackner}, \citenamefont {Brennan}, \citenamefont {Matter}, \citenamefont
  {Park}, \citenamefont {Wright},\ and\ \citenamefont {van~der
  Zwaan}}]{urgenPNAS:12}%
  \BibitemOpen
  \bibfield  {author} {\bibinfo {author} {\bibfnamefont {K.~S.}\ \bibnamefont
  {Lackner}}, \bibinfo {author} {\bibfnamefont {S.}~\bibnamefont {Brennan}},
  \bibinfo {author} {\bibfnamefont {J.M.}\ \bibnamefont {Matter}}, \bibinfo
  {author} {\bibfnamefont {A.-H.~Alissa}\ \bibnamefont {Park}}, \bibinfo
  {author} {\bibfnamefont {A.}~\bibnamefont {Wright}}, \ and\ \bibinfo {author}
  {\bibfnamefont {B.}~\bibnamefont {van~der Zwaan}},\ }\bibfield  {title}
  {\enquote {\bibinfo {title} {The urgency of the development of co2 capture
  from ambient air},}\ }\href@noop {} {\bibfield  {journal} {\bibinfo
  {journal} {Proceedings of the National Academy of Sciences}\ }\textbf
  {\bibinfo {volume} {109}},\ \bibinfo {pages} {13156--13162} (\bibinfo {year}
  {2012})}\BibitemShut {NoStop}%
\bibitem [{\citenamefont {Scott}\ \emph {et~al.}(2013)\citenamefont {Scott},
  \citenamefont {Gilfillan}, \citenamefont {Markusson}, \citenamefont
  {Chalmers},\ and\ \citenamefont {Haszeldine}}]{natclimchange:13}%
  \BibitemOpen
  \bibfield  {author} {\bibinfo {author} {\bibfnamefont {V.}~\bibnamefont
  {Scott}}, \bibinfo {author} {\bibfnamefont {S.}~\bibnamefont {Gilfillan}},
  \bibinfo {author} {\bibfnamefont {N.}~\bibnamefont {Markusson}}, \bibinfo
  {author} {\bibfnamefont {H.}~\bibnamefont {Chalmers}}, \ and\ \bibinfo
  {author} {\bibfnamefont {R.~S.}\ \bibnamefont {Haszeldine}},\ }\bibfield
  {title} {\enquote {\bibinfo {title} {Last chance for carbon capture and
  storage},}\ }\href@noop {} {\bibfield  {journal} {\bibinfo  {journal} {Nature
  Climate Change}\ }\textbf {\bibinfo {volume} {3}},\ \bibinfo {pages}
  {105--111} (\bibinfo {year} {2013})}\BibitemShut {NoStop}%
\bibitem [{\citenamefont {Broom}\ and\ \citenamefont {Thomas}(2013)}]{MRS:13}%
  \BibitemOpen
  \bibfield  {author} {\bibinfo {author} {\bibfnamefont {D.~P.}\ \bibnamefont
  {Broom}}\ and\ \bibinfo {author} {\bibfnamefont {K.~M.}\ \bibnamefont
  {Thomas}},\ }\bibfield  {title} {\enquote {\bibinfo {title} {Gas adsorption
  by nanoporous materials: Future applications and experimental challenges},}\
  }\href@noop {} {\bibfield  {journal} {\bibinfo  {journal} {MRS Bullettin}\
  }\textbf {\bibinfo {volume} {38}},\ \bibinfo {pages} {412--420} (\bibinfo
  {year} {2013})}\BibitemShut {NoStop}%
\bibitem [{\citenamefont {Makal}\ \emph {et~al.}(2012)\citenamefont {Makal},
  \citenamefont {Li}, \citenamefont {Lu},\ and\ \citenamefont
  {Zhou}}]{ChemSocRev:12}%
  \BibitemOpen
  \bibfield  {author} {\bibinfo {author} {\bibfnamefont {T.~A.}\ \bibnamefont
  {Makal}}, \bibinfo {author} {\bibfnamefont {J.~R.}\ \bibnamefont {Li}},
  \bibinfo {author} {\bibfnamefont {W.}~\bibnamefont {Lu}}, \ and\ \bibinfo
  {author} {\bibfnamefont {H.~C.}\ \bibnamefont {Zhou}},\ }\bibfield  {title}
  {\enquote {\bibinfo {title} {Methane storage in advanced porous materials},}\
  }\href@noop {} {\bibfield  {journal} {\bibinfo  {journal} {Chem. Soc. Rev.}\
  }\textbf {\bibinfo {volume} {41}},\ \bibinfo {pages} {7761--7779} (\bibinfo
  {year} {2012})}\BibitemShut {NoStop}%
\bibitem [{\citenamefont {Nandi}\ and\ \citenamefont
  {Uyama}(2014)}]{revco2:14}%
  \BibitemOpen
  \bibfield  {author} {\bibinfo {author} {\bibfnamefont {M.}~\bibnamefont
  {Nandi}}\ and\ \bibinfo {author} {\bibfnamefont {H.}~\bibnamefont {Uyama}},\
  }\bibfield  {title} {\enquote {\bibinfo {title} {Exceptional co2 adsorbing
  materials under different conditions},}\ }\href@noop {} {\bibfield  {journal}
  {\bibinfo  {journal} {Chem. Rec.}\ }\textbf {\bibinfo {volume} {14}},\
  \bibinfo {pages} {1134--1148} (\bibinfo {year} {2014})}\BibitemShut {NoStop}%
\bibitem [{\citenamefont {Gadipelli}\ \emph {et~al.}(2014)\citenamefont
  {Gadipelli}, \citenamefont {Vaiva}, \citenamefont {Zheng-Xiao},\ and\
  \citenamefont {Taner}}]{porouscarb:14}%
  \BibitemOpen
  \bibfield  {author} {\bibinfo {author} {\bibfnamefont {S.}~\bibnamefont
  {Gadipelli}}, \bibinfo {author} {\bibfnamefont {K.}~\bibnamefont {Vaiva}},
  \bibinfo {author} {\bibfnamefont {G.}~\bibnamefont {Zheng-Xiao}}, \ and\
  \bibinfo {author} {\bibfnamefont {Y.}~\bibnamefont {Taner}},\ }\bibfield
  {title} {\enquote {\bibinfo {title} {Exceptional co2 capture in a
  hieararchically porous carbon with simultaneous high surface area and pore
  volume},}\ }\href@noop {} {\bibfield  {journal} {\bibinfo  {journal} {Energy
  Environ. Sci.}\ }\textbf {\bibinfo {volume} {7}},\ \bibinfo {pages}
  {335--342} (\bibinfo {year} {2014})}\BibitemShut {NoStop}%
\bibitem [{\citenamefont {Balasubramanian}\ and\ \citenamefont
  {Chowdhury}(2015)}]{graphadsrev:15}%
  \BibitemOpen
  \bibfield  {author} {\bibinfo {author} {\bibfnamefont {R.}~\bibnamefont
  {Balasubramanian}}\ and\ \bibinfo {author} {\bibfnamefont {S.}~\bibnamefont
  {Chowdhury}},\ }\bibfield  {title} {\enquote {\bibinfo {title} {Recent
  advances and progress in the development of graphene-based adsorbents for co2
  capture},}\ }\href@noop {} {\bibfield  {journal} {\bibinfo  {journal} {J.
  Mater. Chem. A}\ }\textbf {\bibinfo {volume} {3}},\ \bibinfo {pages}
  {21968--21989} (\bibinfo {year} {2015})}\BibitemShut {NoStop}%
\bibitem [{\citenamefont {Burress}\ \emph {et~al.}(2010)\citenamefont
  {Burress}, \citenamefont {Gadipelli}, \citenamefont {Ford}, \citenamefont
  {Simmons}, \citenamefont {Zhou},\ and\ \citenamefont
  {Yildirim}}]{Graphoxi:10}%
  \BibitemOpen
  \bibfield  {author} {\bibinfo {author} {\bibfnamefont {J.~W.}\ \bibnamefont
  {Burress}}, \bibinfo {author} {\bibfnamefont {S.}~\bibnamefont {Gadipelli}},
  \bibinfo {author} {\bibfnamefont {J.}~\bibnamefont {Ford}}, \bibinfo {author}
  {\bibfnamefont {J.~M.}\ \bibnamefont {Simmons}}, \bibinfo {author}
  {\bibfnamefont {W.}~\bibnamefont {Zhou}}, \ and\ \bibinfo {author}
  {\bibfnamefont {T.}~\bibnamefont {Yildirim}},\ }\bibfield  {title} {\enquote
  {\bibinfo {title} {Graphene oxide framework materials: Theoretical
  predictions and experimental results},}\ }\href {\doibase
  10.1002/anie.201003328} {\bibfield  {journal} {\bibinfo  {journal}
  {Angewandte Chemie International Edition}\ }\textbf {\bibinfo {volume}
  {49}},\ \bibinfo {pages} {8902--8904} (\bibinfo {year} {2010})}\BibitemShut
  {NoStop}%
\bibitem [{\citenamefont {Klechikov}\ \emph {et~al.}(2015)\citenamefont
  {Klechikov}, \citenamefont {Mercier}, \citenamefont {Merino}, \citenamefont
  {Blanco}, \citenamefont {Merino},\ and\ \citenamefont
  {Talyzin}}]{carbonnanostruc:15}%
  \BibitemOpen
  \bibfield  {author} {\bibinfo {author} {\bibfnamefont {A.~G.}\ \bibnamefont
  {Klechikov}}, \bibinfo {author} {\bibfnamefont {G.}~\bibnamefont {Mercier}},
  \bibinfo {author} {\bibfnamefont {P.}~\bibnamefont {Merino}}, \bibinfo
  {author} {\bibfnamefont {S.}~\bibnamefont {Blanco}}, \bibinfo {author}
  {\bibfnamefont {C.}~\bibnamefont {Merino}}, \ and\ \bibinfo {author}
  {\bibfnamefont {A.~V.}\ \bibnamefont {Talyzin}},\ }\bibfield  {title}
  {\enquote {\bibinfo {title} {Hydrogen storage in bulk graphene-related
  materials},}\ }\href@noop {} {\bibfield  {journal} {\bibinfo  {journal}
  {Micropor. Mesopor. Mater.}\ }\textbf {\bibinfo {volume} {210}},\ \bibinfo
  {pages} {46--51} (\bibinfo {year} {2015})}\BibitemShut {NoStop}%
\bibitem [{\citenamefont {Ganesan}\ and\ \citenamefont
  {Shaijumon}(2016)}]{activgraph:16}%
  \BibitemOpen
  \bibfield  {author} {\bibinfo {author} {\bibfnamefont {A.}~\bibnamefont
  {Ganesan}}\ and\ \bibinfo {author} {\bibfnamefont {M.~M.}\ \bibnamefont
  {Shaijumon}},\ }\bibfield  {title} {\enquote {\bibinfo {title} {Activated
  graphene-derived porous carbon with exceptional gas adsorption properties},}\
  }\href@noop {} {\bibfield  {journal} {\bibinfo  {journal} {Micropor. Mesopor.
  Mater.}\ }\textbf {\bibinfo {volume} {220}},\ \bibinfo {pages} {21--27}
  (\bibinfo {year} {2016})}\BibitemShut {NoStop}%
\bibitem [{\citenamefont {Bieri}\ \emph {et~al.}(2009)\citenamefont {Bieri},
  \citenamefont {Treier}, \citenamefont {Cai}, \citenamefont {A\"{i}t-Mansour},
  \citenamefont {Ruffieux}, \citenamefont {Gr\"{o}ning}, \citenamefont
  {Gr\"{o}ning}, \citenamefont {Kastler}, \citenamefont {Rieger}, \citenamefont
  {Feng}, \citenamefont {M\"{u}llen},\ and\ \citenamefont {Fasel}}]{Bieri:09}%
  \BibitemOpen
  \bibfield  {author} {\bibinfo {author} {\bibfnamefont {M.}~\bibnamefont
  {Bieri}}, \bibinfo {author} {\bibfnamefont {M.}~\bibnamefont {Treier}},
  \bibinfo {author} {\bibfnamefont {J.}~\bibnamefont {Cai}}, \bibinfo {author}
  {\bibfnamefont {K.}~\bibnamefont {A\"{i}t-Mansour}}, \bibinfo {author}
  {\bibfnamefont {P.}~\bibnamefont {Ruffieux}}, \bibinfo {author}
  {\bibfnamefont {O.}~\bibnamefont {Gr\"{o}ning}}, \bibinfo {author}
  {\bibfnamefont {P.}~\bibnamefont {Gr\"{o}ning}}, \bibinfo {author}
  {\bibfnamefont {M.}~\bibnamefont {Kastler}}, \bibinfo {author} {\bibfnamefont
  {R.}~\bibnamefont {Rieger}}, \bibinfo {author} {\bibfnamefont
  {X.}~\bibnamefont {Feng}}, \bibinfo {author} {\bibfnamefont {K.}~\bibnamefont
  {M\"{u}llen}}, \ and\ \bibinfo {author} {\bibfnamefont {R.}~\bibnamefont
  {Fasel}},\ }\bibfield  {title} {\enquote {\bibinfo {title} {Porous graphenes:
  Two-dimensional polymer synthesis with atomic precision},}\ }\href@noop {}
  {\bibfield  {journal} {\bibinfo  {journal} {Chem. Commun.}\ }\textbf
  {\bibinfo {volume} {45}},\ \bibinfo {pages} {6919--6921} (\bibinfo {year}
  {2009})}\BibitemShut {NoStop}%
\bibitem [{\citenamefont {Li}\ \emph {et~al.}(2010)\citenamefont {Li},
  \citenamefont {Li}, \citenamefont {Liu}, \citenamefont {Guo}, \citenamefont
  {Li},\ and\ \citenamefont {Zhu}}]{graphd-chemcomm:2010}%
  \BibitemOpen
  \bibfield  {author} {\bibinfo {author} {\bibfnamefont {G.}~\bibnamefont
  {Li}}, \bibinfo {author} {\bibfnamefont {Y.}~\bibnamefont {Li}}, \bibinfo
  {author} {\bibfnamefont {H.}~\bibnamefont {Liu}}, \bibinfo {author}
  {\bibfnamefont {Y.}~\bibnamefont {Guo}}, \bibinfo {author} {\bibfnamefont
  {Y.}~\bibnamefont {Li}}, \ and\ \bibinfo {author} {\bibfnamefont {D..}\
  \bibnamefont {Zhu}},\ }\bibfield  {title} {\enquote {\bibinfo {title}
  {Architecture of graphdiyne nanoscale films},}\ }\href@noop {} {\bibfield
  {journal} {\bibinfo  {journal} {Chem. Commun.}\ }\textbf {\bibinfo {volume}
  {46}},\ \bibinfo {pages} {3256--3258} (\bibinfo {year} {2010})}\BibitemShut
  {NoStop}%
\bibitem [{\citenamefont {Narita}\ \emph {et~al.}(1999)\citenamefont {Narita},
  \citenamefont {Nagai}, \citenamefont {Suzuki},\ and\ \citenamefont
  {Nakao}}]{narita:98}%
  \BibitemOpen
  \bibfield  {author} {\bibinfo {author} {\bibfnamefont {N.}~\bibnamefont
  {Narita}}, \bibinfo {author} {\bibfnamefont {S.}~\bibnamefont {Nagai}},
  \bibinfo {author} {\bibfnamefont {S.}~\bibnamefont {Suzuki}}, \ and\ \bibinfo
  {author} {\bibfnamefont {K.}~\bibnamefont {Nakao}},\ }\bibfield  {title}
  {\enquote {\bibinfo {title} {Optimized geometries and electronic structures
  of graphyne and its family},}\ }\href@noop {} {\bibfield  {journal} {\bibinfo
   {journal} {Phys. Rev. B}\ }\textbf {\bibinfo {volume} {58}},\ \bibinfo
  {pages} {11009--11014} (\bibinfo {year} {1999})}\BibitemShut {NoStop}%
\bibitem [{\citenamefont {Li}\ \emph {et~al.}(2014)\citenamefont {Li},
  \citenamefont {Xu}, \citenamefont {Liu},\ and\ \citenamefont
  {Li}}]{graphd_review:2014}%
  \BibitemOpen
  \bibfield  {author} {\bibinfo {author} {\bibfnamefont {Y.}~\bibnamefont
  {Li}}, \bibinfo {author} {\bibfnamefont {L.}~\bibnamefont {Xu}}, \bibinfo
  {author} {\bibfnamefont {H.}~\bibnamefont {Liu}}, \ and\ \bibinfo {author}
  {\bibfnamefont {Y.}~\bibnamefont {Li}},\ }\bibfield  {title} {\enquote
  {\bibinfo {title} {Graphdiyne and graphyne: from theoretical predictions to
  practical construction},}\ }\href@noop {} {\bibfield  {journal} {\bibinfo
  {journal} {Chem. Soc. Rev.}\ }\textbf {\bibinfo {volume} {43}},\ \bibinfo
  {pages} {2572--2586} (\bibinfo {year} {2014})}\BibitemShut {NoStop}%
\bibitem [{\citenamefont {Zhou}\ \emph {et~al.}(2015)\citenamefont {Zhou},
  \citenamefont {Gao}, \citenamefont {Liu}, \citenamefont {Xie}, \citenamefont
  {Yang}, \citenamefont {Zhang}, \citenamefont {Zhang}, \citenamefont {Liu},
  \citenamefont {Li}, \citenamefont {Zhang},\ and\ \citenamefont
  {Liu}}]{graphdsynt:15}%
  \BibitemOpen
  \bibfield  {author} {\bibinfo {author} {\bibfnamefont {Jingyuan}\
  \bibnamefont {Zhou}}, \bibinfo {author} {\bibfnamefont {Xin}\ \bibnamefont
  {Gao}}, \bibinfo {author} {\bibfnamefont {Rong}\ \bibnamefont {Liu}},
  \bibinfo {author} {\bibfnamefont {Ziqian}\ \bibnamefont {Xie}}, \bibinfo
  {author} {\bibfnamefont {Jin}\ \bibnamefont {Yang}}, \bibinfo {author}
  {\bibfnamefont {Shuqing}\ \bibnamefont {Zhang}}, \bibinfo {author}
  {\bibfnamefont {Gengmin}\ \bibnamefont {Zhang}}, \bibinfo {author}
  {\bibfnamefont {Huibiao}\ \bibnamefont {Liu}}, \bibinfo {author}
  {\bibfnamefont {Yuliang}\ \bibnamefont {Li}}, \bibinfo {author}
  {\bibfnamefont {Jin}\ \bibnamefont {Zhang}}, \ and\ \bibinfo {author}
  {\bibfnamefont {Zhongfan}\ \bibnamefont {Liu}},\ }\bibfield  {title}
  {\enquote {\bibinfo {title} {Synthesis of graphdiyne nanowalls using
  acetylenic coupling reaction},}\ }\href@noop {} {\bibfield  {journal}
  {\bibinfo  {journal} {Journal of the American Chemical Society}\ }\textbf
  {\bibinfo {volume} {137}},\ \bibinfo {pages} {7596--7599} (\bibinfo {year}
  {2015})}\BibitemShut {NoStop}%
\bibitem [{\citenamefont {Cranford}\ and\ \citenamefont
  {Buehler}(2012)}]{nanoscalemitbis:2012}%
  \BibitemOpen
  \bibfield  {author} {\bibinfo {author} {\bibfnamefont {S.~W.}\ \bibnamefont
  {Cranford}}\ and\ \bibinfo {author} {\bibfnamefont {M.~J.}\ \bibnamefont
  {Buehler}},\ }\bibfield  {title} {\enquote {\bibinfo {title} {Selective
  hydrogen purification through graphdiyne under ambient temperature and
  pressure},}\ }\href@noop {} {\bibfield  {journal} {\bibinfo  {journal}
  {Nanoscale}\ }\textbf {\bibinfo {volume} {4}},\ \bibinfo {pages} {4587--4593}
  (\bibinfo {year} {2012})}\BibitemShut {NoStop}%
\bibitem [{\citenamefont {Lin}\ and\ \citenamefont
  {Buehler}(2013)}]{nanoscalemit:2013}%
  \BibitemOpen
  \bibfield  {author} {\bibinfo {author} {\bibfnamefont {S.}~\bibnamefont
  {Lin}}\ and\ \bibinfo {author} {\bibfnamefont {M.~J.}\ \bibnamefont
  {Buehler}},\ }\bibfield  {title} {\enquote {\bibinfo {title} {Mechanics and
  molecular filtration performance of graphyne nanoweb membranes for selective
  water purification},}\ }\href@noop {} {\bibfield  {journal} {\bibinfo
  {journal} {Nanoscale}\ }\textbf {\bibinfo {volume} {5}},\ \bibinfo {pages}
  {11801--11807} (\bibinfo {year} {2013})}\BibitemShut {NoStop}%
\bibitem [{\citenamefont {Bartolomei}\ \emph
  {et~al.}(2014{\natexlab{a}})\citenamefont {Bartolomei}, \citenamefont
  {Carmona-Novillo}, \citenamefont {Hern{\'a}ndez}, \citenamefont
  {Campos-Mart{\'\i}nez}, \citenamefont {Pirani}, \citenamefont {Giorgi},\ and\
  \citenamefont {Yamashita}}]{jpclours:2014}%
  \BibitemOpen
  \bibfield  {author} {\bibinfo {author} {\bibfnamefont {M.}~\bibnamefont
  {Bartolomei}}, \bibinfo {author} {\bibfnamefont {E.}~\bibnamefont
  {Carmona-Novillo}}, \bibinfo {author} {\bibfnamefont {M.~I.}\ \bibnamefont
  {Hern{\'a}ndez}}, \bibinfo {author} {\bibfnamefont {J.}~\bibnamefont
  {Campos-Mart{\'\i}nez}}, \bibinfo {author} {\bibfnamefont {F.}~\bibnamefont
  {Pirani}}, \bibinfo {author} {\bibfnamefont {G.}~\bibnamefont {Giorgi}}, \
  and\ \bibinfo {author} {\bibfnamefont {K.}~\bibnamefont {Yamashita}},\
  }\bibfield  {title} {\enquote {\bibinfo {title} {Penetration barrier of water
  through graphynes' pores: First-principles predictions and force field
  optimization},}\ }\href@noop {} {\bibfield  {journal} {\bibinfo  {journal}
  {J. Phys. Chem. Lett.}\ }\textbf {\bibinfo {volume} {5}},\ \bibinfo {pages}
  {751--755} (\bibinfo {year} {2014}{\natexlab{a}})}\BibitemShut {NoStop}%
\bibitem [{\citenamefont {Bartolomei}\ \emph
  {et~al.}(2014{\natexlab{b}})\citenamefont {Bartolomei}, \citenamefont
  {Carmona-Novillo}, \citenamefont {Hern{\'a}ndez}, \citenamefont
  {Campos-Mart{\'\i}nez}, \citenamefont {Pirani},\ and\ \citenamefont
  {Giorgi}}]{jpccours:2014}%
  \BibitemOpen
  \bibfield  {author} {\bibinfo {author} {\bibfnamefont {M.}~\bibnamefont
  {Bartolomei}}, \bibinfo {author} {\bibfnamefont {E.}~\bibnamefont
  {Carmona-Novillo}}, \bibinfo {author} {\bibfnamefont {M.~I.}\ \bibnamefont
  {Hern{\'a}ndez}}, \bibinfo {author} {\bibfnamefont {J.}~\bibnamefont
  {Campos-Mart{\'\i}nez}}, \bibinfo {author} {\bibfnamefont {F.}~\bibnamefont
  {Pirani}}, \ and\ \bibinfo {author} {\bibfnamefont {G.}~\bibnamefont
  {Giorgi}},\ }\bibfield  {title} {\enquote {\bibinfo {title} {Graphdiyne
  pores: "ad hoc" openings for helium separation applications},}\ }\href@noop
  {} {\bibfield  {journal} {\bibinfo  {journal} {J. Phys. Chem. C}\ }\textbf
  {\bibinfo {volume} {118}},\ \bibinfo {pages} {29966--29972} (\bibinfo {year}
  {2014}{\natexlab{b}})}\BibitemShut {NoStop}%
\bibitem [{\citenamefont {Hern{\'a}ndez}\ \emph {et~al.}(2015)\citenamefont
  {Hern{\'a}ndez}, \citenamefont {Bartolomei},\ and\ \citenamefont
  {Campos-Mart{\'\i}nez}}]{heliumisotop:15}%
  \BibitemOpen
  \bibfield  {author} {\bibinfo {author} {\bibfnamefont {M.~I.}\ \bibnamefont
  {Hern{\'a}ndez}}, \bibinfo {author} {\bibfnamefont {M.}~\bibnamefont
  {Bartolomei}}, \ and\ \bibinfo {author} {\bibfnamefont {J.}~\bibnamefont
  {Campos-Mart{\'\i}nez}},\ }\bibfield  {title} {\enquote {\bibinfo {title}
  {Transmission of helium isotopes through graphdiyne pores: Tunneling versus
  zero point energy effects},}\ }\href@noop {} {\bibfield  {journal} {\bibinfo
  {journal} {The Journal of Physical Chemistry A}\ }\textbf {\bibinfo {volume}
  {119}},\ \bibinfo {pages} {10743--10749} (\bibinfo {year}
  {2015})}\BibitemShut {NoStop}%
\bibitem [{\citenamefont {Zheng}\ \emph {et~al.}(2012)\citenamefont {Zheng},
  \citenamefont {Luo}, \citenamefont {Liu}, \citenamefont {Quhe}, \citenamefont
  {Zheng}, \citenamefont {Tang}, \citenamefont {Gao}, \citenamefont {Nagase},\
  and\ \citenamefont {Lu}}]{bitri:2012}%
  \BibitemOpen
  \bibfield  {author} {\bibinfo {author} {\bibfnamefont {Q.}~\bibnamefont
  {Zheng}}, \bibinfo {author} {\bibfnamefont {G.}~\bibnamefont {Luo}}, \bibinfo
  {author} {\bibfnamefont {Q.}~\bibnamefont {Liu}}, \bibinfo {author}
  {\bibfnamefont {R.}~\bibnamefont {Quhe}}, \bibinfo {author} {\bibfnamefont
  {J.}~\bibnamefont {Zheng}}, \bibinfo {author} {\bibfnamefont
  {K.}~\bibnamefont {Tang}}, \bibinfo {author} {\bibfnamefont {Z.}~\bibnamefont
  {Gao}}, \bibinfo {author} {\bibfnamefont {S.}~\bibnamefont {Nagase}}, \ and\
  \bibinfo {author} {\bibfnamefont {J.}~\bibnamefont {Lu}},\ }\bibfield
  {title} {\enquote {\bibinfo {title} {Structural and electronic properties of
  bilayer and trilayer graphdiyne},}\ }\href@noop {} {\bibfield  {journal}
  {\bibinfo  {journal} {Nanoscale}\ }\textbf {\bibinfo {volume} {4}},\ \bibinfo
  {pages} {3990--3996} (\bibinfo {year} {2012})}\BibitemShut {NoStop}%
\bibitem [{\citenamefont {Dec{\'e}r{\'e}}\ \emph {et~al.}(2013)\citenamefont
  {Dec{\'e}r{\'e}}, \citenamefont {Lepetit},\ and\ \citenamefont
  {Chauvin}}]{bulk-graphyne:2013}%
  \BibitemOpen
  \bibfield  {author} {\bibinfo {author} {\bibfnamefont {J.M.}\ \bibnamefont
  {Dec{\'e}r{\'e}}}, \bibinfo {author} {\bibfnamefont {C.}~\bibnamefont
  {Lepetit}}, \ and\ \bibinfo {author} {\bibfnamefont {R.}~\bibnamefont
  {Chauvin}},\ }\bibfield  {title} {\enquote {\bibinfo {title} {Carbo-graphite:
  Structural, mechanical, and electronic properties},}\ }\href@noop {}
  {\bibfield  {journal} {\bibinfo  {journal} {J. Phys. Chem. C}\ }\textbf
  {\bibinfo {volume} {117}},\ \bibinfo {pages} {21671--21681} (\bibinfo {year}
  {2013})}\BibitemShut {NoStop}%
\bibitem [{\citenamefont {Luo}\ \emph {et~al.}(2013)\citenamefont {Luo},
  \citenamefont {Zheng}, \citenamefont {Mei}, \citenamefont {Lu},\ and\
  \citenamefont {Nagase}}]{bulk-graphd:2013}%
  \BibitemOpen
  \bibfield  {author} {\bibinfo {author} {\bibfnamefont {G.}~\bibnamefont
  {Luo}}, \bibinfo {author} {\bibfnamefont {Q.}~\bibnamefont {Zheng}}, \bibinfo
  {author} {\bibfnamefont {W.}~\bibnamefont {Mei}}, \bibinfo {author}
  {\bibfnamefont {J.}~\bibnamefont {Lu}}, \ and\ \bibinfo {author}
  {\bibfnamefont {S.}~\bibnamefont {Nagase}},\ }\bibfield  {title} {\enquote
  {\bibinfo {title} {Structural, electronic, and optical properties of bulk
  graphdiyne},}\ }\href@noop {} {\bibfield  {journal} {\bibinfo  {journal} {J.
  Phys. Chem. C}\ }\textbf {\bibinfo {volume} {117}},\ \bibinfo {pages}
  {13072--13079} (\bibinfo {year} {2013})}\BibitemShut {NoStop}%
\bibitem [{\citenamefont {Bartolomei}\ \emph {et~al.}(2005)\citenamefont
  {Bartolomei}, \citenamefont {Carmona-Novillo},\ and\ \citenamefont
  {Giorgi}}]{carbonour:2015}%
  \BibitemOpen
  \bibfield  {author} {\bibinfo {author} {\bibfnamefont {M.}~\bibnamefont
  {Bartolomei}}, \bibinfo {author} {\bibfnamefont {E.}~\bibnamefont
  {Carmona-Novillo}}, \ and\ \bibinfo {author} {\bibfnamefont {G.}~\bibnamefont
  {Giorgi}},\ }\bibfield  {title} {\enquote {\bibinfo {title} {First principles
  investigation of hydrogen physical adsorption on graphynes' layers},}\
  }\href@noop {} {\bibfield  {journal} {\bibinfo  {journal} {Carbon}\ }\textbf
  {\bibinfo {volume} {95}},\ \bibinfo {pages} {1076--1081} (\bibinfo {year}
  {2005})}\BibitemShut {NoStop}%
\bibitem [{\citenamefont {Vidali}\ \emph {et~al.}(1991)\citenamefont {Vidali},
  \citenamefont {Ihm}, \citenamefont {Kim},\ and\ \citenamefont
  {Cole}}]{Vidali:91}%
  \BibitemOpen
  \bibfield  {author} {\bibinfo {author} {\bibfnamefont {G.}~\bibnamefont
  {Vidali}}, \bibinfo {author} {\bibfnamefont {G.}~\bibnamefont {Ihm}},
  \bibinfo {author} {\bibfnamefont {H.~Y.}\ \bibnamefont {Kim}}, \ and\
  \bibinfo {author} {\bibfnamefont {M.~W.}\ \bibnamefont {Cole}},\ }\bibfield
  {title} {\enquote {\bibinfo {title} {Potentials of physical adsorption},}\
  }\href@noop {} {\bibfield  {journal} {\bibinfo  {journal} {Surf. Sci. Rep.}\
  }\textbf {\bibinfo {volume} {12}},\ \bibinfo {pages} {133--181} (\bibinfo
  {year} {1991})}\BibitemShut {NoStop}%
\bibitem [{\citenamefont {Rubes}\ \emph {et~al.}(2010)\citenamefont {Rubes},
  \citenamefont {Kysilka}, \citenamefont {Nachtigall},\ and\ \citenamefont
  {Bludsky}}]{dft-cc:10}%
  \BibitemOpen
  \bibfield  {author} {\bibinfo {author} {\bibfnamefont {M.}~\bibnamefont
  {Rubes}}, \bibinfo {author} {\bibfnamefont {J.}~\bibnamefont {Kysilka}},
  \bibinfo {author} {\bibfnamefont {P.}~\bibnamefont {Nachtigall}}, \ and\
  \bibinfo {author} {\bibfnamefont {O.}~\bibnamefont {Bludsky}},\ }\bibfield
  {title} {\enquote {\bibinfo {title} {Dft/cc investigation of physical
  adsorption on a graphite (0001) surface},}\ }\href@noop {} {\bibfield
  {journal} {\bibinfo  {journal} {Phys. Chem. Chem. Phys.}\ }\textbf {\bibinfo
  {volume} {12}},\ \bibinfo {pages} {6438--6444} (\bibinfo {year}
  {2010})}\BibitemShut {NoStop}%
\bibitem [{\citenamefont {Silvestrelli}\ and\ \citenamefont
  {Ambrosetti}(2014)}]{dft-vdw:14}%
  \BibitemOpen
  \bibfield  {author} {\bibinfo {author} {\bibfnamefont {P.~L.}\ \bibnamefont
  {Silvestrelli}}\ and\ \bibinfo {author} {\bibfnamefont {A.}~\bibnamefont
  {Ambrosetti}},\ }\bibfield  {title} {\enquote {\bibinfo {title} {Including
  screening in van der waals corrected density functional theory calculations:
  The case of atoms and small molecules physisorbed on graphene},}\ }\href@noop
  {} {\bibfield  {journal} {\bibinfo  {journal} {J. Chem. Phys.}\ }\textbf
  {\bibinfo {volume} {140}},\ \bibinfo {pages} {124107} (\bibinfo {year}
  {2014})}\BibitemShut {NoStop}%
\bibitem [{\citenamefont {Kresse}\ and\ \citenamefont
  {Furthm\"{u}ller}(1996)}]{VASP:96}%
  \BibitemOpen
  \bibfield  {author} {\bibinfo {author} {\bibfnamefont {G.}~\bibnamefont
  {Kresse}}\ and\ \bibinfo {author} {\bibfnamefont {J.}~\bibnamefont
  {Furthm\"{u}ller}},\ }\bibfield  {title} {\enquote {\bibinfo {title}
  {Efficient iterative schemes for ab initio total-energy calculations using a
  plane-wave basis set},}\ }\href@noop {} {\bibfield  {journal} {\bibinfo
  {journal} {Phys. Rev. B}\ }\textbf {\bibinfo {volume} {54}},\ \bibinfo
  {pages} {11169--11186} (\bibinfo {year} {1996})}\BibitemShut {NoStop}%
\bibitem [{\citenamefont {Perdew}\ \emph {et~al.}(1996)\citenamefont {Perdew},
  \citenamefont {Burke},\ and\ \citenamefont {Ernzerhof}}]{pbe:96}%
  \BibitemOpen
  \bibfield  {author} {\bibinfo {author} {\bibfnamefont {J.P.}\ \bibnamefont
  {Perdew}}, \bibinfo {author} {\bibfnamefont {K.}~\bibnamefont {Burke}}, \
  and\ \bibinfo {author} {\bibfnamefont {M.}~\bibnamefont {Ernzerhof}},\
  }\bibfield  {title} {\enquote {\bibinfo {title} {Generalized gradient
  approximation made simple},}\ }\href@noop {} {\bibfield  {journal} {\bibinfo
  {journal} {Phys. Rev. Lett.}\ }\textbf {\bibinfo {volume} {77}},\ \bibinfo
  {pages} {3865--3868} (\bibinfo {year} {1996})}\BibitemShut {NoStop}%
\bibitem [{\citenamefont {Bl\"{o}chl}(1994)}]{PAW1}%
  \BibitemOpen
  \bibfield  {author} {\bibinfo {author} {\bibfnamefont {P.E.}\ \bibnamefont
  {Bl\"{o}chl}},\ }\bibfield  {title} {\enquote {\bibinfo {title} {Projector
  augmented-wave method},}\ }\href@noop {} {\bibfield  {journal} {\bibinfo
  {journal} {Phys. Rev. B}\ }\textbf {\bibinfo {volume} {50}},\ \bibinfo
  {pages} {17953--17979} (\bibinfo {year} {1994})}\BibitemShut {NoStop}%
\bibitem [{\citenamefont {Kresse}\ and\ \citenamefont {Joubert}(1999)}]{PAW2}%
  \BibitemOpen
  \bibfield  {author} {\bibinfo {author} {\bibfnamefont {G.}~\bibnamefont
  {Kresse}}\ and\ \bibinfo {author} {\bibfnamefont {D.}~\bibnamefont
  {Joubert}},\ }\bibfield  {title} {\enquote {\bibinfo {title} {From ultrasoft
  pseudopotentials to the projector augmented-wave method},}\ }\href@noop {}
  {\bibfield  {journal} {\bibinfo  {journal} {Phys. Rev. B}\ }\textbf {\bibinfo
  {volume} {59}},\ \bibinfo {pages} {1758--1775} (\bibinfo {year}
  {1999})}\BibitemShut {NoStop}%
\bibitem [{\citenamefont {Grimme}\ \emph {et~al.}(2011)\citenamefont {Grimme},
  \citenamefont {Ehrlich},\ and\ \citenamefont {Goerigk}}]{Grimme:11}%
  \BibitemOpen
  \bibfield  {author} {\bibinfo {author} {\bibfnamefont {S.}~\bibnamefont
  {Grimme}}, \bibinfo {author} {\bibfnamefont {S.}~\bibnamefont {Ehrlich}}, \
  and\ \bibinfo {author} {\bibfnamefont {L.}~\bibnamefont {Goerigk}},\
  }\bibfield  {title} {\enquote {\bibinfo {title} {Effect of the damping
  function in dispersion corrected density functional theory},}\ }\href@noop {}
  {\bibfield  {journal} {\bibinfo  {journal} {J. Comput. Chem.}\ }\textbf
  {\bibinfo {volume} {32}},\ \bibinfo {pages} {1456--1465} (\bibinfo {year}
  {2011})}\BibitemShut {NoStop}%
\bibitem [{\citenamefont {Brandenburg}\ \emph {et~al.}(2014)\citenamefont
  {Brandenburg}, \citenamefont {Hochheim}, \citenamefont {Bredow},\ and\
  \citenamefont {Grimme}}]{Grimme:14}%
  \BibitemOpen
  \bibfield  {author} {\bibinfo {author} {\bibfnamefont {J.~G.}\ \bibnamefont
  {Brandenburg}}, \bibinfo {author} {\bibfnamefont {M.}~\bibnamefont
  {Hochheim}}, \bibinfo {author} {\bibfnamefont {T.}~\bibnamefont {Bredow}}, \
  and\ \bibinfo {author} {\bibfnamefont {S.}~\bibnamefont {Grimme}},\
  }\bibfield  {title} {\enquote {\bibinfo {title} {Low-cost quantum chemical
  methods for noncovalent interactions},}\ }\href@noop {} {\bibfield  {journal}
  {\bibinfo  {journal} {J. Phys. Chem. Lett.}\ }\textbf {\bibinfo {volume}
  {5}},\ \bibinfo {pages} {4275--4284} (\bibinfo {year} {2014})}\BibitemShut
  {NoStop}%
\bibitem [{\citenamefont {Grimme}\ \emph {et~al.}(2010)\citenamefont {Grimme},
  \citenamefont {Antony}, \citenamefont {Ehrlich},\ and\ \citenamefont
  {Krieg}}]{Grimme:10}%
  \BibitemOpen
  \bibfield  {author} {\bibinfo {author} {\bibfnamefont {S.}~\bibnamefont
  {Grimme}}, \bibinfo {author} {\bibfnamefont {J.}~\bibnamefont {Antony}},
  \bibinfo {author} {\bibfnamefont {S.}~\bibnamefont {Ehrlich}}, \ and\
  \bibinfo {author} {\bibfnamefont {H.}~\bibnamefont {Krieg}},\ }\bibfield
  {title} {\enquote {\bibinfo {title} {A consistent and accurate ab initio
  parametrization of density functional dispersion correction (dft-d) for the
  94 elements h-pu},}\ }\href@noop {} {\bibfield  {journal} {\bibinfo
  {journal} {J. Chem. Phys.}\ }\textbf {\bibinfo {volume} {132}},\ \bibinfo
  {pages} {154104} (\bibinfo {year} {2010})}\BibitemShut {NoStop}%
\bibitem [{\citenamefont {Piton\'ak}\ and\ \citenamefont
  {Hesselmann}(2010)}]{mp2c}%
  \BibitemOpen
  \bibfield  {author} {\bibinfo {author} {\bibfnamefont {M.}~\bibnamefont
  {Piton\'ak}}\ and\ \bibinfo {author} {\bibfnamefont {A.}~\bibnamefont
  {Hesselmann}},\ }\bibfield  {title} {\enquote {\bibinfo {title} {Accurate
  intermolecular interaction energies from a combination of mp2 and tddft
  response theory},}\ }\href@noop {} {\bibfield  {journal} {\bibinfo  {journal}
  {J. Chem. Theory Comput.}\ }\textbf {\bibinfo {volume} {6012}},\ \bibinfo
  {pages} {168--178} (\bibinfo {year} {2010})}\BibitemShut {NoStop}%
\bibitem [{\citenamefont {Werner}\ \emph {et~al.}(2012)\citenamefont {Werner},
  \citenamefont {Knowles}, \citenamefont {Lindh}, \citenamefont {Manby},
  \citenamefont {{Sch\"{u}tz}}, \citenamefont {Celani}, \citenamefont {Korona},
  \citenamefont {Rauhut}, \citenamefont {Amos}, \citenamefont {Bernhardsson},
  \citenamefont {Berning}, \citenamefont {Cooper}, \citenamefont {Deegan},
  \citenamefont {Dobbyn}, \citenamefont {Eckert}, \citenamefont {Hampel},
  \citenamefont {Hetzer}, \citenamefont {Lloyd}, \citenamefont {{McNicholas}},
  \citenamefont {Meyer}, \citenamefont {Mura}, \citenamefont {Nicklass},
  \citenamefont {Palmieri}, \citenamefont {Pitzer}, \citenamefont {Schumann},
  \citenamefont {Stoll}, \citenamefont {Stone}, \citenamefont {Tarroni},\ and\
  \citenamefont {Thorsteinsson}}]{MOLPRO}%
  \BibitemOpen
  \bibfield  {author} {\bibinfo {author} {\bibfnamefont {H.-J.}\ \bibnamefont
  {Werner}}, \bibinfo {author} {\bibfnamefont {P.~J.}\ \bibnamefont {Knowles}},
  \bibinfo {author} {\bibfnamefont {R.}~\bibnamefont {Lindh}}, \bibinfo
  {author} {\bibfnamefont {F.~R.}\ \bibnamefont {Manby}}, \bibinfo {author}
  {\bibfnamefont {M.}~\bibnamefont {{Sch\"{u}tz}}}, \bibinfo {author}
  {\bibfnamefont {P.}~\bibnamefont {Celani}}, \bibinfo {author} {\bibfnamefont
  {T.}~\bibnamefont {Korona}}, \bibinfo {author} {\bibfnamefont
  {G.}~\bibnamefont {Rauhut}}, \bibinfo {author} {\bibfnamefont {R.~D.}\
  \bibnamefont {Amos}}, \bibinfo {author} {\bibfnamefont {A.}~\bibnamefont
  {Bernhardsson}}, \bibinfo {author} {\bibfnamefont {A.}~\bibnamefont
  {Berning}}, \bibinfo {author} {\bibfnamefont {D.~L.}\ \bibnamefont {Cooper}},
  \bibinfo {author} {\bibfnamefont {M.~J.~O.}\ \bibnamefont {Deegan}}, \bibinfo
  {author} {\bibfnamefont {A.~J.}\ \bibnamefont {Dobbyn}}, \bibinfo {author}
  {\bibfnamefont {F.}~\bibnamefont {Eckert}}, \bibinfo {author} {\bibfnamefont
  {C.}~\bibnamefont {Hampel}}, \bibinfo {author} {\bibfnamefont
  {G.}~\bibnamefont {Hetzer}}, \bibinfo {author} {\bibfnamefont {A.~W.}\
  \bibnamefont {Lloyd}}, \bibinfo {author} {\bibfnamefont {S.~J.}\ \bibnamefont
  {{McNicholas}}}, \bibinfo {author} {\bibfnamefont {W.}~\bibnamefont {Meyer}},
  \bibinfo {author} {\bibfnamefont {M.~E.}\ \bibnamefont {Mura}}, \bibinfo
  {author} {\bibfnamefont {A.}~\bibnamefont {Nicklass}}, \bibinfo {author}
  {\bibfnamefont {P.}~\bibnamefont {Palmieri}}, \bibinfo {author}
  {\bibfnamefont {R.}~\bibnamefont {Pitzer}}, \bibinfo {author} {\bibfnamefont
  {U.}~\bibnamefont {Schumann}}, \bibinfo {author} {\bibfnamefont
  {H.}~\bibnamefont {Stoll}}, \bibinfo {author} {\bibfnamefont {A.~J.}\
  \bibnamefont {Stone}}, \bibinfo {author} {\bibfnamefont {R.}~\bibnamefont
  {Tarroni}}, \ and\ \bibinfo {author} {\bibfnamefont {T.}~\bibnamefont
  {Thorsteinsson}},\ }\href@noop {} {\enquote {\bibinfo {title} {Molpro,
  version2012.1, a package of ab initio programs},}\ } (\bibinfo {year}
  {2012}),\ \bibinfo {note} {seehttp://www.molpro.net}\BibitemShut {NoStop}%
\bibitem [{\citenamefont {Hesselmann}\ and\ \citenamefont
  {Korona}(2011)}]{mp2c-full}%
  \BibitemOpen
  \bibfield  {author} {\bibinfo {author} {\bibfnamefont {A.}~\bibnamefont
  {Hesselmann}}\ and\ \bibinfo {author} {\bibfnamefont {T.}~\bibnamefont
  {Korona}},\ }\bibfield  {title} {\enquote {\bibinfo {title} {On the accuracy
  of dft-sapt, mp2, scs-mp2, mp2c, and dft+disp methods for the interaction
  energies of endohedral complexes of the c(60) fullerene with a rare gas
  atom},}\ }\href@noop {} {\bibfield  {journal} {\bibinfo  {journal} {Phys.
  Chem. Chem. Phys.}\ }\textbf {\bibinfo {volume} {13}},\ \bibinfo {pages}
  {732--743} (\bibinfo {year} {2011})}\BibitemShut {NoStop}%
\bibitem [{\citenamefont {Bartolomei}\ \emph {et~al.}(2013)\citenamefont
  {Bartolomei}, \citenamefont {Carmona-Novillo}, \citenamefont {Hern{\'a}ndez},
  \citenamefont {Campos-Mart{\'\i}nez},\ and\ \citenamefont
  {Pirani}}]{grapheneours:2013}%
  \BibitemOpen
  \bibfield  {author} {\bibinfo {author} {\bibfnamefont {M.}~\bibnamefont
  {Bartolomei}}, \bibinfo {author} {\bibfnamefont {E.}~\bibnamefont
  {Carmona-Novillo}}, \bibinfo {author} {\bibfnamefont {M.~I.}\ \bibnamefont
  {Hern{\'a}ndez}}, \bibinfo {author} {\bibfnamefont {J.}~\bibnamefont
  {Campos-Mart{\'\i}nez}}, \ and\ \bibinfo {author} {\bibfnamefont
  {F.}~\bibnamefont {Pirani}},\ }\bibfield  {title} {\enquote {\bibinfo {title}
  {Global potentials for the interaction between rare gases and graphene-based
  surfaces: An atom-bond pairwise additive representation},}\ }\href@noop {}
  {\bibfield  {journal} {\bibinfo  {journal} {J. Phys. Chem. C}\ }\textbf
  {\bibinfo {volume} {117}},\ \bibinfo {pages} {10512--10522} (\bibinfo {year}
  {2013})}\BibitemShut {NoStop}%
\bibitem [{\citenamefont {Pei}(2012)}]{mech-graphdiyne:2012}%
  \BibitemOpen
  \bibfield  {author} {\bibinfo {author} {\bibfnamefont {Y.}~\bibnamefont
  {Pei}},\ }\bibfield  {title} {\enquote {\bibinfo {title} {Mechanical
  properties of graphdiyne sheet},}\ }\href@noop {} {\bibfield  {journal}
  {\bibinfo  {journal} {Physica B}\ }\textbf {\bibinfo {volume} {407}},\
  \bibinfo {pages} {4436--4439} (\bibinfo {year} {2012})}\BibitemShut {NoStop}%
\bibitem [{\citenamefont {Kendall}\ \emph {et~al.}(1992)\citenamefont
  {Kendall}, \citenamefont {Dunning},\ and\ \citenamefont
  {Harrison}}]{Dunning}%
  \BibitemOpen
  \bibfield  {author} {\bibinfo {author} {\bibfnamefont {R.~A.}\ \bibnamefont
  {Kendall}}, \bibinfo {author} {\bibfnamefont {T.~H.}\ \bibnamefont
  {Dunning}}, \ and\ \bibinfo {author} {\bibfnamefont {R.~J.}\ \bibnamefont
  {Harrison}},\ }\bibfield  {title} {\enquote {\bibinfo {title} {Electron
  affinities of the first-row atoms revisited. systematic basis sets and wave
  functions},}\ }\href@noop {} {\bibfield  {journal} {\bibinfo  {journal} {J.
  Chem. Phys.}\ }\textbf {\bibinfo {volume} {96}},\ \bibinfo {pages}
  {6796--6806} (\bibinfo {year} {1992})}\BibitemShut {NoStop}%
\bibitem [{\citenamefont {Boys}\ and\ \citenamefont
  {Bernardi}(1970)}]{Boys:70}%
  \BibitemOpen
  \bibfield  {author} {\bibinfo {author} {\bibfnamefont {S.F.}\ \bibnamefont
  {Boys}}\ and\ \bibinfo {author} {\bibfnamefont {F.}~\bibnamefont
  {Bernardi}},\ }\bibfield  {title} {\enquote {\bibinfo {title} {The
  calculation of small molecular interactions by the differences of separate
  total energies. some procedures with reduced errors},}\ }\href@noop {}
  {\bibfield  {journal} {\bibinfo  {journal} {Mol. Phys.}\ }\textbf {\bibinfo
  {volume} {19}},\ \bibinfo {pages} {553--566} (\bibinfo {year}
  {1970})}\BibitemShut {NoStop}%
\bibitem [{\citenamefont {Frisch}\ \emph {et~al.}()\citenamefont {Frisch},
  \citenamefont {Trucks}, \citenamefont {Schlegel}, \citenamefont {Scuseria},
  \citenamefont {Robb}, \citenamefont {Cheeseman}, \citenamefont {Scalmani},
  \citenamefont {Barone}, \citenamefont {Mennucci}, \citenamefont {Petersson},
  \citenamefont {Nakatsuji}, \citenamefont {Caricato}, \citenamefont {Li},
  \citenamefont {Hratchian}, \citenamefont {Izmaylov}, \citenamefont {Bloino},
  \citenamefont {Zheng}, \citenamefont {Sonnenberg}, \citenamefont {Hada},
  \citenamefont {Ehara}, \citenamefont {Toyota}, \citenamefont {Fukuda},
  \citenamefont {Hasegawa}, \citenamefont {Ishida}, \citenamefont {Nakajima},
  \citenamefont {Honda}, \citenamefont {Kitao}, \citenamefont {Nakai},
  \citenamefont {Vreven}, \citenamefont {Montgomery}, \citenamefont {Peralta},
  \citenamefont {Ogliaro}, \citenamefont {Bearpark}, \citenamefont {Heyd},
  \citenamefont {Brothers}, \citenamefont {Kudin}, \citenamefont {Staroverov},
  \citenamefont {Kobayashi}, \citenamefont {Normand}, \citenamefont
  {Raghavachari}, \citenamefont {Rendell}, \citenamefont {Burant},
  \citenamefont {Iyengar}, \citenamefont {Tomasi}, \citenamefont {Cossi},
  \citenamefont {Rega}, \citenamefont {Millam}, \citenamefont {Klene},
  \citenamefont {Knox}, \citenamefont {Cross}, \citenamefont {Bakken},
  \citenamefont {Adamo}, \citenamefont {Jaramillo}, \citenamefont {Gomperts},
  \citenamefont {Stratmann}, \citenamefont {Yazyev}, \citenamefont {Austin},
  \citenamefont {Cammi}, \citenamefont {Pomelli}, \citenamefont {Ochterski},
  \citenamefont {Martin}, \citenamefont {Morokuma}, \citenamefont {Zakrzewski},
  \citenamefont {Voth}, \citenamefont {Salvador}, \citenamefont {Dannenberg},
  \citenamefont {Dapprich}, \citenamefont {Daniels}, \citenamefont {Farkas},
  \citenamefont {Foresman}, \citenamefont {Ortiz}, \citenamefont {Cioslowski},\
  and\ \citenamefont {Fox}}]{g09}%
  \BibitemOpen
  \bibfield  {author} {\bibinfo {author} {\bibfnamefont {M.~J.}\ \bibnamefont
  {Frisch}}, \bibinfo {author} {\bibfnamefont {G.~W.}\ \bibnamefont {Trucks}},
  \bibinfo {author} {\bibfnamefont {H.~B.}\ \bibnamefont {Schlegel}}, \bibinfo
  {author} {\bibfnamefont {G.~E.}\ \bibnamefont {Scuseria}}, \bibinfo {author}
  {\bibfnamefont {M.~A.}\ \bibnamefont {Robb}}, \bibinfo {author}
  {\bibfnamefont {J.~R.}\ \bibnamefont {Cheeseman}}, \bibinfo {author}
  {\bibfnamefont {G.}~\bibnamefont {Scalmani}}, \bibinfo {author}
  {\bibfnamefont {V.}~\bibnamefont {Barone}}, \bibinfo {author} {\bibfnamefont
  {B.}~\bibnamefont {Mennucci}}, \bibinfo {author} {\bibfnamefont {G.~A.}\
  \bibnamefont {Petersson}}, \bibinfo {author} {\bibfnamefont {H.}~\bibnamefont
  {Nakatsuji}}, \bibinfo {author} {\bibfnamefont {M.}~\bibnamefont {Caricato}},
  \bibinfo {author} {\bibfnamefont {X.}~\bibnamefont {Li}}, \bibinfo {author}
  {\bibfnamefont {H.~P.}\ \bibnamefont {Hratchian}}, \bibinfo {author}
  {\bibfnamefont {A.~F.}\ \bibnamefont {Izmaylov}}, \bibinfo {author}
  {\bibfnamefont {J.}~\bibnamefont {Bloino}}, \bibinfo {author} {\bibfnamefont
  {G.}~\bibnamefont {Zheng}}, \bibinfo {author} {\bibfnamefont {J.~L.}\
  \bibnamefont {Sonnenberg}}, \bibinfo {author} {\bibfnamefont
  {M.}~\bibnamefont {Hada}}, \bibinfo {author} {\bibfnamefont {M.}~\bibnamefont
  {Ehara}}, \bibinfo {author} {\bibfnamefont {K.}~\bibnamefont {Toyota}},
  \bibinfo {author} {\bibfnamefont {R.}~\bibnamefont {Fukuda}}, \bibinfo
  {author} {\bibfnamefont {J.}~\bibnamefont {Hasegawa}}, \bibinfo {author}
  {\bibfnamefont {M.}~\bibnamefont {Ishida}}, \bibinfo {author} {\bibfnamefont
  {T.}~\bibnamefont {Nakajima}}, \bibinfo {author} {\bibfnamefont
  {Y.}~\bibnamefont {Honda}}, \bibinfo {author} {\bibfnamefont
  {O.}~\bibnamefont {Kitao}}, \bibinfo {author} {\bibfnamefont
  {H.}~\bibnamefont {Nakai}}, \bibinfo {author} {\bibfnamefont
  {T.}~\bibnamefont {Vreven}}, \bibinfo {author} {\bibfnamefont {J.~A.}\
  \bibnamefont {Montgomery}, \bibfnamefont {{Jr.}}}, \bibinfo {author}
  {\bibfnamefont {J.~E.}\ \bibnamefont {Peralta}}, \bibinfo {author}
  {\bibfnamefont {F.}~\bibnamefont {Ogliaro}}, \bibinfo {author} {\bibfnamefont
  {M.}~\bibnamefont {Bearpark}}, \bibinfo {author} {\bibfnamefont {J.~J.}\
  \bibnamefont {Heyd}}, \bibinfo {author} {\bibfnamefont {E.}~\bibnamefont
  {Brothers}}, \bibinfo {author} {\bibfnamefont {K.~N.}\ \bibnamefont {Kudin}},
  \bibinfo {author} {\bibfnamefont {V.~N.}\ \bibnamefont {Staroverov}},
  \bibinfo {author} {\bibfnamefont {R.}~\bibnamefont {Kobayashi}}, \bibinfo
  {author} {\bibfnamefont {J.}~\bibnamefont {Normand}}, \bibinfo {author}
  {\bibfnamefont {K.}~\bibnamefont {Raghavachari}}, \bibinfo {author}
  {\bibfnamefont {A.}~\bibnamefont {Rendell}}, \bibinfo {author} {\bibfnamefont
  {J.~C.}\ \bibnamefont {Burant}}, \bibinfo {author} {\bibfnamefont {S.~S.}\
  \bibnamefont {Iyengar}}, \bibinfo {author} {\bibfnamefont {J.}~\bibnamefont
  {Tomasi}}, \bibinfo {author} {\bibfnamefont {M.}~\bibnamefont {Cossi}},
  \bibinfo {author} {\bibfnamefont {N.}~\bibnamefont {Rega}}, \bibinfo {author}
  {\bibfnamefont {J.~M.}\ \bibnamefont {Millam}}, \bibinfo {author}
  {\bibfnamefont {M.}~\bibnamefont {Klene}}, \bibinfo {author} {\bibfnamefont
  {J.~E.}\ \bibnamefont {Knox}}, \bibinfo {author} {\bibfnamefont {J.~B.}\
  \bibnamefont {Cross}}, \bibinfo {author} {\bibfnamefont {V.}~\bibnamefont
  {Bakken}}, \bibinfo {author} {\bibfnamefont {C.}~\bibnamefont {Adamo}},
  \bibinfo {author} {\bibfnamefont {J.}~\bibnamefont {Jaramillo}}, \bibinfo
  {author} {\bibfnamefont {R.}~\bibnamefont {Gomperts}}, \bibinfo {author}
  {\bibfnamefont {R.~E.}\ \bibnamefont {Stratmann}}, \bibinfo {author}
  {\bibfnamefont {O.}~\bibnamefont {Yazyev}}, \bibinfo {author} {\bibfnamefont
  {A.~J.}\ \bibnamefont {Austin}}, \bibinfo {author} {\bibfnamefont
  {R.}~\bibnamefont {Cammi}}, \bibinfo {author} {\bibfnamefont
  {C.}~\bibnamefont {Pomelli}}, \bibinfo {author} {\bibfnamefont {J.~W.}\
  \bibnamefont {Ochterski}}, \bibinfo {author} {\bibfnamefont {R.~L.}\
  \bibnamefont {Martin}}, \bibinfo {author} {\bibfnamefont {K.}~\bibnamefont
  {Morokuma}}, \bibinfo {author} {\bibfnamefont {V.~G.}\ \bibnamefont
  {Zakrzewski}}, \bibinfo {author} {\bibfnamefont {G.~A.}\ \bibnamefont
  {Voth}}, \bibinfo {author} {\bibfnamefont {P.}~\bibnamefont {Salvador}},
  \bibinfo {author} {\bibfnamefont {J.~J.}\ \bibnamefont {Dannenberg}},
  \bibinfo {author} {\bibfnamefont {S.}~\bibnamefont {Dapprich}}, \bibinfo
  {author} {\bibfnamefont {A.~D.}\ \bibnamefont {Daniels}}, \bibinfo {author}
  {\bibfnamefont {\"{O}}\ \bibnamefont {Farkas}}, \bibinfo {author}
  {\bibfnamefont {J.~B.}\ \bibnamefont {Foresman}}, \bibinfo {author}
  {\bibfnamefont {J.~V.}\ \bibnamefont {Ortiz}}, \bibinfo {author}
  {\bibfnamefont {J.}~\bibnamefont {Cioslowski}}, \ and\ \bibinfo {author}
  {\bibfnamefont {D.~J.}\ \bibnamefont {Fox}},\ }\href@noop {} {\enquote
  {\bibinfo {title} {Gaussian\,09 {R}evision {E}.01},}\ }\bibinfo {note}
  {Gaussian Inc. Wallingford CT 2009}\BibitemShut {NoStop}%
\bibitem [{\citenamefont {Brandenburg}\ \emph {et~al.}(2013)\citenamefont
  {Brandenburg}, \citenamefont {Alessio}, \citenamefont {Civalleri},
  \citenamefont {Peintinger}, \citenamefont {Bredow},\ and\ \citenamefont
  {Grimme}}]{Brandenburg:2013}%
  \BibitemOpen
  \bibfield  {author} {\bibinfo {author} {\bibfnamefont {J.~G.}\ \bibnamefont
  {Brandenburg}}, \bibinfo {author} {\bibfnamefont {M.}~\bibnamefont
  {Alessio}}, \bibinfo {author} {\bibfnamefont {B.}~\bibnamefont {Civalleri}},
  \bibinfo {author} {\bibfnamefont {M.~F.}\ \bibnamefont {Peintinger}},
  \bibinfo {author} {\bibfnamefont {T.}~\bibnamefont {Bredow}}, \ and\ \bibinfo
  {author} {\bibfnamefont {S.}~\bibnamefont {Grimme}},\ }\bibfield  {title}
  {\enquote {\bibinfo {title} {Geometrical correction for the inter- and
  intramolecular basis set superposition error in periodic density functional
  theory calculations},}\ }\href@noop {} {\bibfield  {journal} {\bibinfo
  {journal} {J. Phys. Chem. A}\ }\textbf {\bibinfo {volume} {117}},\ \bibinfo
  {pages} {9282--9292} (\bibinfo {year} {2013})}\BibitemShut {NoStop}%
\bibitem [{\citenamefont {Denbigh}(1940)}]{Denbigh}%
  \BibitemOpen
  \bibfield  {author} {\bibinfo {author} {\bibfnamefont {K.~G.}\ \bibnamefont
  {Denbigh}},\ }\bibfield  {title} {\enquote {\bibinfo {title} {The
  polarisabilities of bonds - i},}\ }\href@noop {} {\bibfield  {journal}
  {\bibinfo  {journal} {Trans. Faraday Soc.}\ }\textbf {\bibinfo {volume}
  {36}},\ \bibinfo {pages} {0936--0947} (\bibinfo {year} {1940})}\BibitemShut
  {NoStop}%
\bibitem [{\citenamefont {Presser}\ \emph {et~al.}(2011)\citenamefont
  {Presser}, \citenamefont {McDonough}, \citenamefont {Yeon},\ and\
  \citenamefont {Gogotsi}}]{CO2porediamter}%
  \BibitemOpen
  \bibfield  {author} {\bibinfo {author} {\bibfnamefont {V.}~\bibnamefont
  {Presser}}, \bibinfo {author} {\bibfnamefont {J.}~\bibnamefont {McDonough}},
  \bibinfo {author} {\bibfnamefont {S.~H.}\ \bibnamefont {Yeon}}, \ and\
  \bibinfo {author} {\bibfnamefont {Y.}~\bibnamefont {Gogotsi}},\ }\bibfield
  {title} {\enquote {\bibinfo {title} {Effect of pore size on carbon dioxide
  sorption by carbide derived carbon},}\ }\href@noop {} {\bibfield  {journal}
  {\bibinfo  {journal} {Energy Environ. Sci.}\ }\textbf {\bibinfo {volume}
  {4}},\ \bibinfo {pages} {3059--3066} (\bibinfo {year} {2011})}\BibitemShut
  {NoStop}%
\end{thebibliography}

\newpage

\begin{table}
 \caption{Binding energies, zero point vibrational energy corrections and
   adsorption enthalpies for a single molecule hosted within an inner pore of
   the porous graphite (see Fig. 4). All values are in meV.}
 \label{tab1}
 \begin{tabular}{cccc}
 \hline
  & Binding energy$^a$&  ZPVE$^b$  & $\Delta$$_{ads}$H \\
\hline
CO$_2$ & 312.0 & 12.7 & -299.3 \\
H$_2$O & 224.3  & 32.4 & -191.8 \\
N$_2$ & 203.0  & 14.3 & -188.7 \\
H$_2$ & 96.4   & 31.6 & -64.8 \\
\hline 
\end{tabular}
\\
$^a$ From MP2C computations. $^b$ From harmonic frequency calculations at the
DFT level of theory assuming the pores as rigid structures. 
\end{table}

\newpage

\begin{figure}[h]
\includegraphics[width=12.cm,angle=-90.]{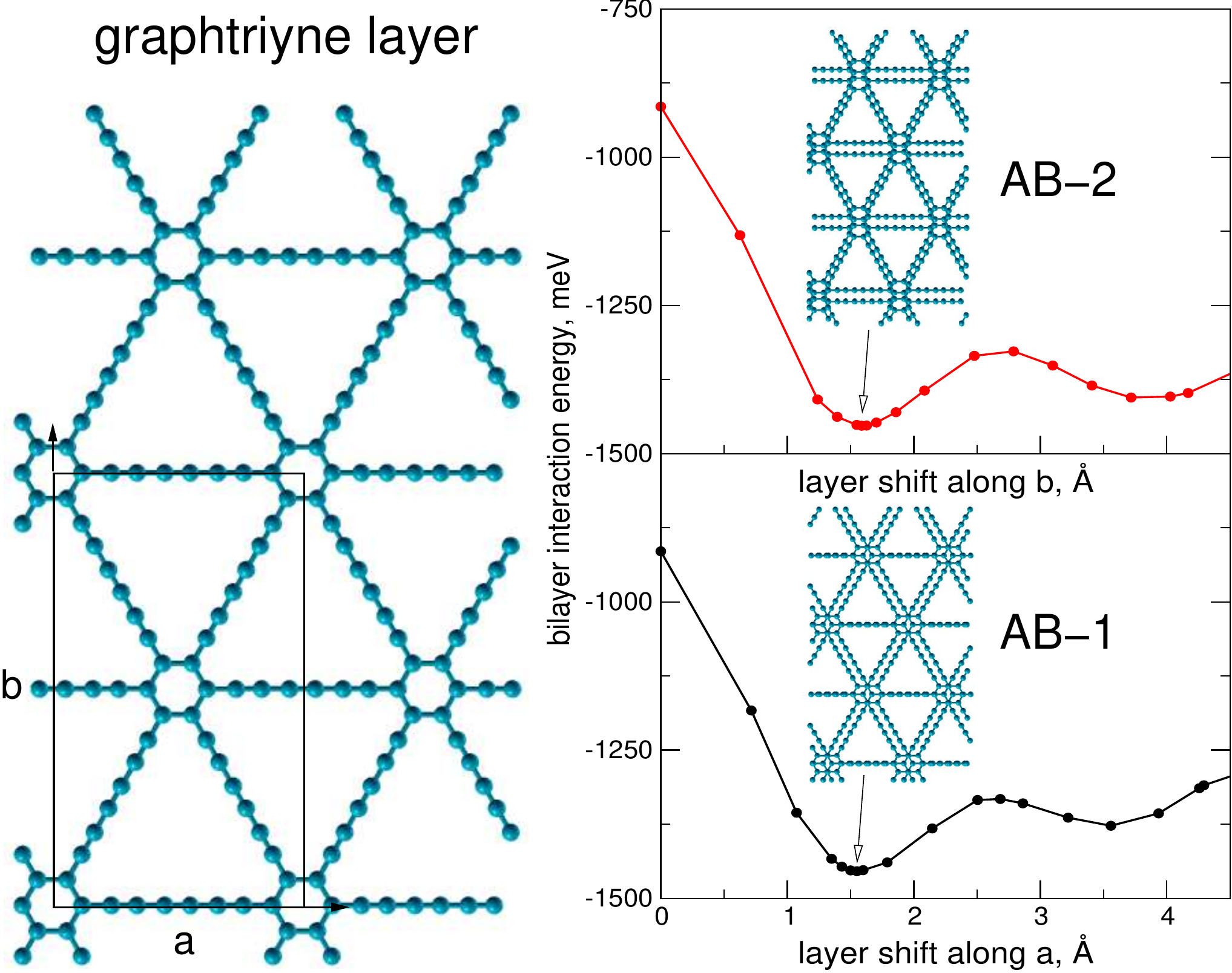}
\caption[]{ Left part: Top view of the  supercell describing the graphtriyne layer used 
 in the periodic DFT calculations together with {\it a} and {\it b}
 lattice parameters.
 Right part: bilayer interaction energy (relative to the
 asymptotic layer separation) as a function of the shift of one layer upon the 
 other along the {\it a} (lower panel) and {\it b} (upper panel) directions. 
The interlayer distance has been kept fixed at 3.45 \AA\, and the shift is
measured with respect to the least favourable AA stacking, that is with carbon
atoms on both layers having the same coordinates along {\it a} and  {\it b}.
The most favourable bilayer structures are also reported in each panel and
labelled as AB-1 and AB-2.}
\label{fig1}
\end{figure}

\begin{figure}[h]
\includegraphics[width=11.cm,angle=-90.]{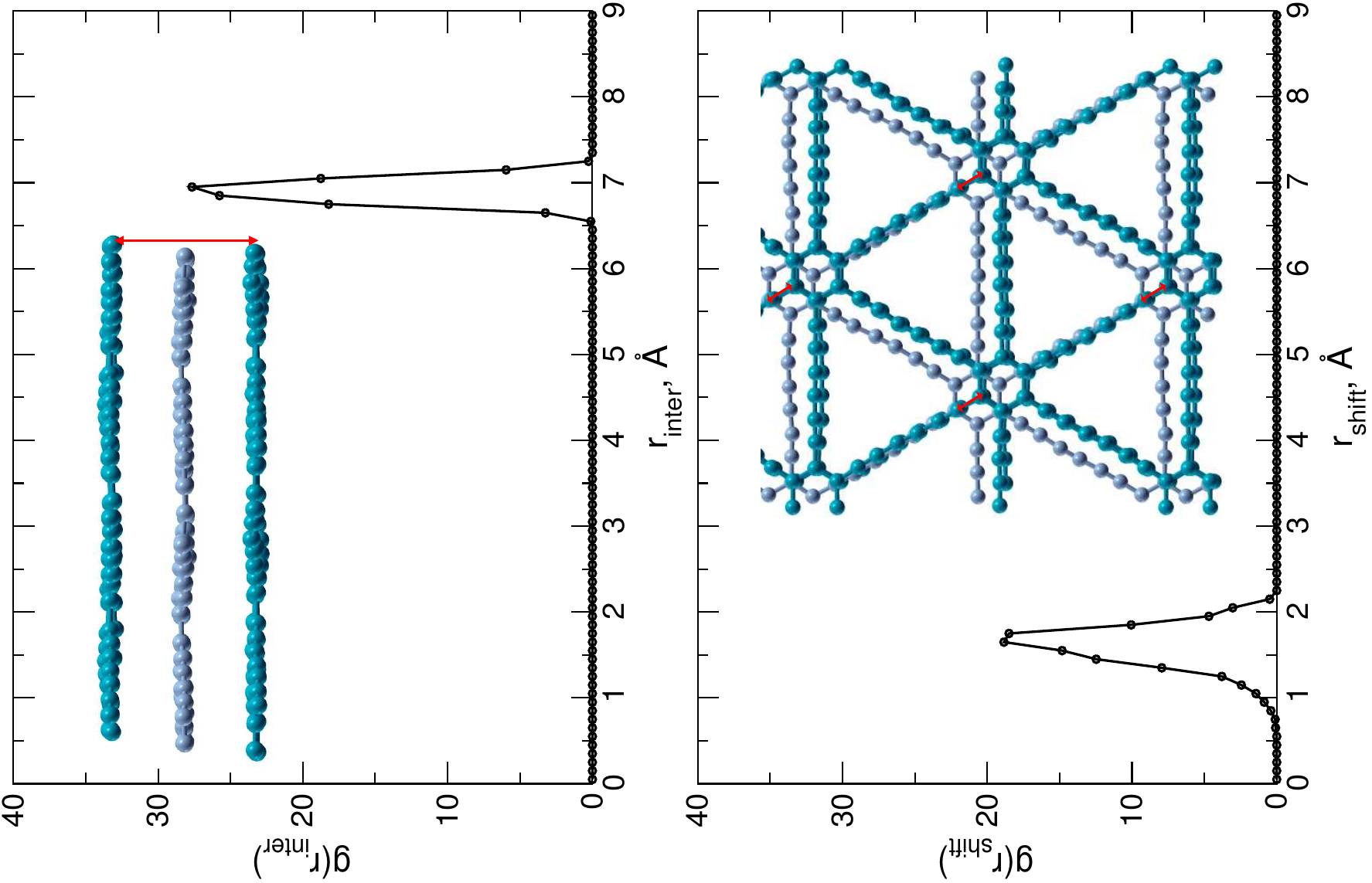}
\caption[]{ Graphtriyne trilayer: radial distributions of the interlayer distance between the outer
  layers (upper panel) and of the inner layer displacement (lower panel) with
  respect to the external ones. Side and top views of the structure obtained
  after molecular dynamics equilibration are also reported 
 in the upper and lower panel, respectively. The inner layer is
  shown in grey for the sake of a better visibility.
 The red double-headed arrows provide graphical representations of the
 interlayer distance and inner layer shift.}
\label{fig2}
\end{figure}

\begin{figure}[h]
\includegraphics[width=11.cm,angle=-90.]{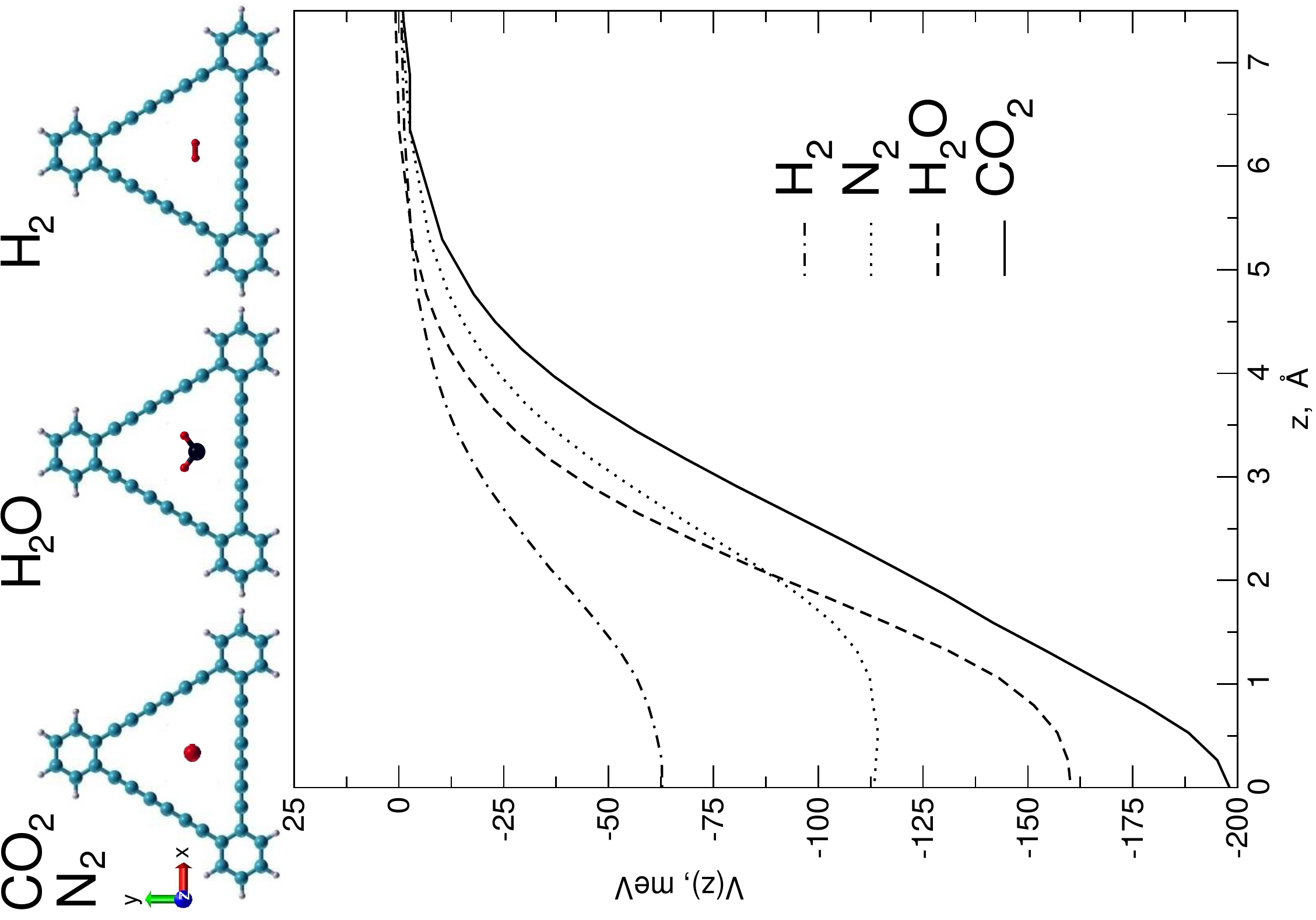}
\caption[]{ Upper part: CO$_2$, N$_2$, H$_2$O and H$_2$ optimal in-pore configuration 
 for their interaction with a graphtriyne pore represented by its smallest
 molecular precursor. CO$_2$ and N$_2$ are in a perpendicular geometry with
 respect to the planar opening while H$_2$O and H$_2$ lie in a coplanar
 arrangement.
 Lower part: adsorption energy profiles for the specific molecule obtained at the MP2C level of theory
  as a function of $z$, the distance from the molecule center of mass to the geometric center of the pore.
Throughout the calculation the relative configuration of the molecule-pore
complex is kept frozen as in the corresponding one shown in the upper part.
}
\label{fig3}
\end{figure}

\newpage

\begin{figure*}[t]
\includegraphics[width=11.5cm,angle=-90.]{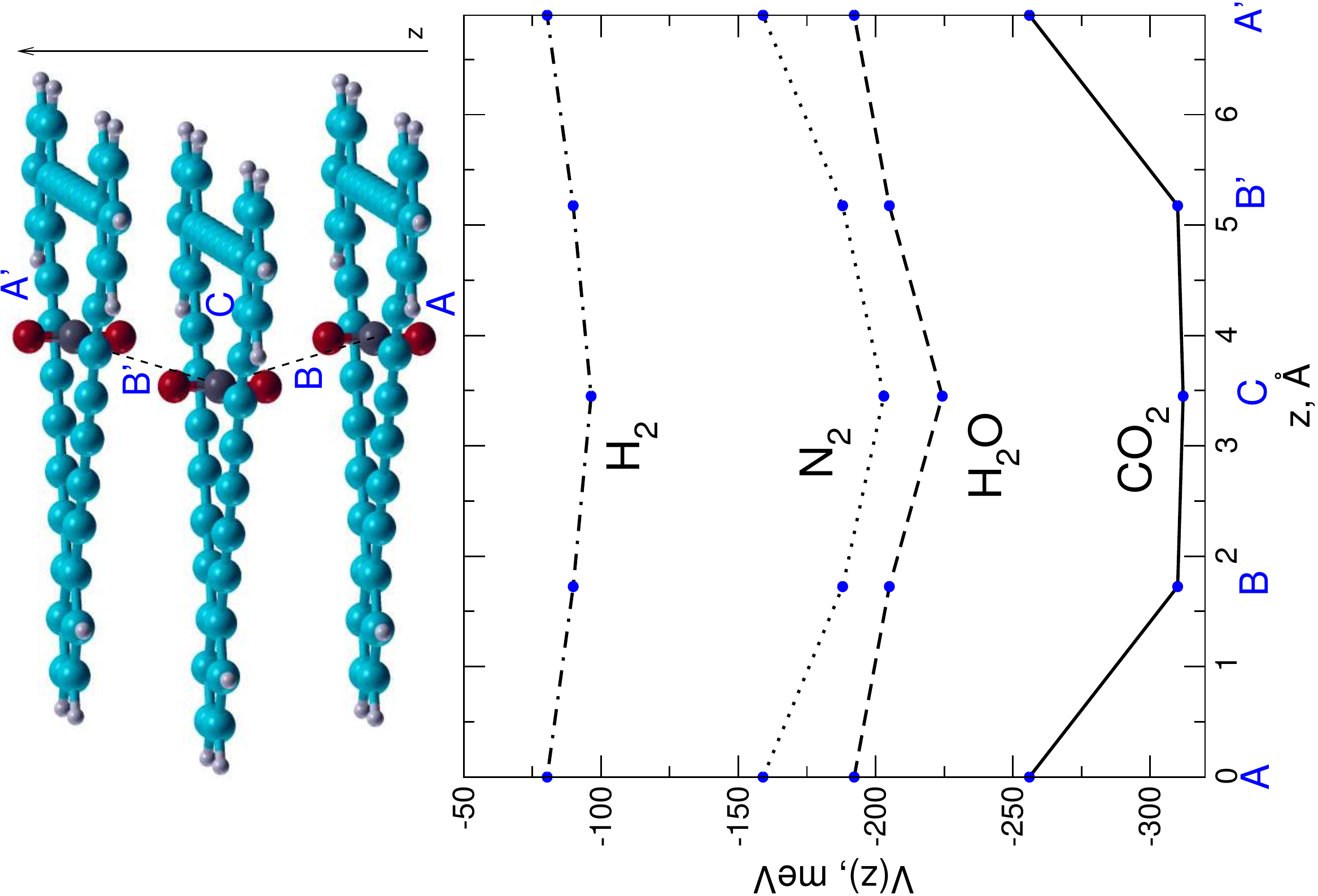}
\caption[]{ Interaction energy evolution of the CO$_2$,  H$_2$O, N$_2$, and H$_2$ molecules crossing the nanoporous
  graphite. 
 A prototype consisting of three parallel
  graphtriyne pores in ABA-1  arrangement is used.  The A, A', B, B' and C letters indicate
  different adsorption sites within the multi-layer: A, A' and C locations correspond to in-pore configurations
 while B and B' sites to intercalation equivalent positions. The
 five adsorption sites lie on the dashed lines which join the
 geometric centers of adjacent pores; the layers separation is fixed at
 3.45 \AA\, while the displacement of the inner pore is fixed at 1.6 \AA, the
 geometry parameters provided by
 the calculations shown in Fig. 2. }
\label{fig4}
\end{figure*}

\clearpage
\newpage







%

\end{document}